\documentclass[aps,prd,twocolumn,nofootinbib,floatfix,superscriptaddress]{revtex4-2}

\usepackage{graphicx}
\usepackage{amsmath,amssymb}
\usepackage{bm}
\usepackage{xcolor}
\usepackage[hidelinks]{hyperref} 
\usepackage{float} 
\usepackage{orcidlink}
\usepackage{placeins}

\usepackage{enumitem}
\usepackage{booktabs}

\newcommand{\NLA}{NLA}
\usepackage{aas_macros}

\begin{document}

\title{Baryonic feedback and intrinsic alignments in profile: a frequentist view of DES Y3 cosmic shear}

\author{Anoma~Ganguly\,\orcidlink{0000-0002-5915-4245}}
\affiliation{Department of Astronomy, University of Arizona, Tucson, AZ 85721, USA}

\author{Elisabeth Krause\,\orcidlink{0000-0001-8356-2014}}
\affiliation{Department of Astronomy, University of Arizona, Tucson, AZ 85721, USA}
\affiliation{Department of Physics, University of Arizona, Tucson, AZ 85721, USA}

\author{Tanvi Karwal\,\orcidlink{0000-0002-1384-9949}}
\affiliation{Department of Astronomy and Astrophysics, University of Chicago, 5640 South Ellis Avenue, Chicago, IL, 60637, USA}
\affiliation{Kavli Institute for Cosmological Physics, Enrico Fermi Institute, University of Chicago, 5640 South Ellis Avenue, Chicago, IL, 60637, USA}

\author{Vivian Miranda\,\orcidlink{0000-0003-4776-0333}}
\affiliation{C. N. Yang Institute for Theoretical Physics, Stony Brook University, Stony Brook, NY 11794, USA}

\begin{abstract}
Current cosmic shear surveys primarily constrain $S_8$ and $\Omega_\mathrm{m}$, while most other model parameters remain dominated by priors.
Consequently, marginal posteriors are vulnerable to prior-volume effects, particularly through degeneracies among $S_8$, baryonic feedback, and intrinsic alignment (IA) model parameters.
To isolate data-driven constraints from prior volume effects, we compare profile likelihoods with Bayesian posteriors for DES Y3 cosmic shear under the DES Y3+KiDS-1000 Hybrid model.
For cosmic shear alone, the profile global best fit and the marginal posterior mean agree in $S_8$, but the profile likelihood reveals a strong degeneracy between feedback strength and $S_8$, leaving the upper $1\sigma$ limit on $S_8$ determined by the assumed feedback range. External CMB information on parameters other than $\Omega_\mathrm{m}$ and $S_8$ breaks this degeneracy, suppressing prior-volume effects and reconciling the profiles with the posteriors. With CMB data included, small-scale cosmic shear data prefers moderate baryonic feedback strengths consistent with hydrodynamical simulations calibrated to observed gas fractions.
Furthermore, additional unconstrained parameters in the tidal alignment and tidal torquing (TATT) model result in weaker constraints, shifting and broadening the $S_8$ profile relative to the nonlinear linear alignment (NLA) model. Adding DES Y3 shear ratios mitigates this degeneracy, bringing profile likelihood and marginal posteriors into close agreement.
\end{abstract}

\maketitle

\section{Introduction}
\label{sec:intro}
Bayesian inference remains the standard framework for cosmological analyses, constraining the parameters of interest while marginalizing over the many additional parameters needed to model astrophysical and observational systematics. Marginalization, however, is vulnerable to prior-volume or projection effects. When weakly constrained nuisance parameters are degenerate with the parameters of interest, the choice of priors can displace the peak of the marginal posterior away from the maximum-likelihood estimate, producing apparent tensions that may not reflect the constraining power of the data~\cite{GomezValent2022, Herold2025}.
 
Profile likelihood offers a complementary frequentist approach that replaces marginalization with optimization. At each fixed value of the parameter of interest, the likelihood is maximized over all remaining parameters, yielding the best achievable data fit independently of prior volume. Profile likelihoods have been applied to several cosmological analyses. The first systematic application to \textit{Planck} data found close agreement with the Bayesian $\Lambda$CDM constraints and used the Feldman--Cousins prescription for the neutrino mass bound~\cite{Planck2014PL, FeldmanCousins1998}, an approach since revisited with newer data including DESI~\cite{NaredoTuero2024, Chebat2026}. Profile likelihoods have also shown that Bayesian constraints are shaped by the prior in analyses of the effective number of neutrino species~\cite{Hamann2012}, early dark energy~\cite{Herold2022EDE, Herold2023}, decaying dark matter~\cite{Holm2023DDM}, EFTofLSS full-shape clustering~\cite{Holm2023}, and $w_0 w_a$CDM fits to DESI full-shape data~\cite{Morawetz2025}. This comparison also serves as a robustness check for dynamical dark energy, showing strong agreement between the two methodologies when applied to DESI, CMB, and supernova datasets~\cite{HeroldKarwal2025}.
 
Cosmic shear, the weak gravitational lensing of background galaxies by large-scale structure, is particularly susceptible to prior-volume effects because it constrains few combinations of its model parameters, primarily $S_8 \equiv \sigma_8 (\Omega_\mathrm{m}/0.3)^{0.5}$, while most other model parameters remain prior dominated. Stage-III surveys, including the Dark Energy Survey (DES)~\cite{Amon2022, Secco2022}, the Kilo-Degree Survey (KiDS)~\cite{Asgari2021}, and the Hyper Suprime-Cam (HSC)~\cite{Li2023, Dalal2023}, have generally found values of $S_8$ below the prediction from \textit{Planck} CMB measurements~\cite{Planck2018}. More recent results are mixed. A joint analysis of DES Y3 and KiDS-1000 with a common ``Hybrid'' model is consistent with \textit{Planck} at the $1.7\sigma$ level~\cite{DESKiDS2023}, and the final KiDS-Legacy analysis agrees with \textit{Planck}, a change attributed primarily to improved redshift calibration~\cite{Wright2025}. DES Year~6, in contrast, continues to find a lower $S_8$~\cite{DESY6shear}.
 
Because the data constrain little beyond $S_8$, the marginal posteriors depend on the prior volume of the remaining parameters. In the Hybrid analysis, the marginal mean $S_8 = 0.790^{+0.018}_{-0.014}$ lies below the maximum a posteriori (MAP) value of $0.801$, a difference the authors attribute to posterior skewness and to projection of the asymmetric neutrino mass prior; with $\Sigma m_\nu$ fixed, the two agree~\cite{DESKiDS2023}. Recent stage-III analyses therefore also
report the MAP, which is noisy to estimate from chains, and projected joint highest posterior density
(PJ-HPD) intervals~\cite{Joachimi2021, Asgari2021, Amon2022, Secco2022, DESKiDS2023,DESY6shear}. These measures reduce the problem within the Bayesian framework, whereas a profile likelihood removes the dependence on prior volume altogether (Sec.~\ref{sec:stats} and references therein).
 
Baryonic feedback is a prominent example of such a degeneracy. Stronger suppression of the small-scale
matter power spectrum allows a higher $S_8$ for the same lensing signal~\cite{AmonEfstathiou2022, Preston2023};
see Ref.~\cite{Chisari2019} for a review of feedback effects on the matter power spectrum. Marginalizing over
feedback with halo-model or principal-component parameterizations removes the resulting biases at the cost of
weaker constraints~\cite{Semboloni2013, Mead2015, Eifler2015, Huang2019}. It also introduces projection
effects, which in simulated LSST analyses are dominated by baryonic feedback~\cite{Robertson2026}. The Hybrid model varies $\log_{10}(T_\mathrm{AGN}/\mathrm{K})$ over $[7.3, 8.0]$, and replacing it with a
dark-matter-only power spectrum lowers $S_8$ by $0.4\sigma$~\cite{DESKiDS2023}. In DES Year~6, $S_8$
increases as the upper prior bound on $\log_{10}(T_\mathrm{AGN}/\mathrm{K})$ is raised~\cite{DESY6shear}. The
marginal $S_8$ may therefore depend on where the prior on $T_\mathrm{AGN}$ ends.

Analyses that constrain the feedback strength with small scales or external data find that the inferred
$S_8$ follows the preferred feedback strength. Including small scales in DES Y3 with a flexible baryon model
raises $S_8$ by $0.2\sigma$~\cite{Arico2023}. Combining DES Y3 with kinetic Sunyaev--Zel'dovich measurements
prefers stronger feedback and a higher $S_8$~\cite{Bigwood2024}, and calibrating the feedback with X-ray and
kSZ data raises $S_8$ from $0.803$ to $0.832$, compared with scale cuts~\cite{Bigwood2025}. KiDS-Legacy and
HSC Year~3, in contrast, find little evidence for strong feedback. In KiDS-Legacy, fixing the cosmology to
\textit{Planck} gives $\log_{10}(T_\mathrm{AGN}/\mathrm{K}) \leq 7.71$ at $68\%$ confidence~\cite{Wright2025},
and HSC Year~3 small scales allow at most $5\%$ suppression of the matter power spectrum~\cite{Terasawa2025}.%
\footnote{Reanalyses of DES data that model baryons with principal components from hydrodynamical simulations
find weak to moderate feedback and an $S_8$ that depends little on the baryon prior~\cite{Huang2021, Xu2024,
Xu2025}.} The BAHAMAS simulations reproduce the observed halo gas fractions for $\log_{10}(T_\mathrm{AGN}/\mathrm{K})$ between $7.6$ and $8.0$~\cite{McCarthy2017, Mead2021}.
 
The choice of intrinsic alignment (IA) model also shifts $S_8$. Replacing the fiducial nonlinear
alignment (NLA) model of the Hybrid analysis with the tidal alignment and tidal torquing (TATT) model
lowers $S_8$ by $0.9\sigma$ and increases its uncertainty by $35\%$~\cite{DESKiDS2023}. The MAP values
of the two models differ by $1.3\sigma$ as well, so this shift is not only a projection effect
\cite{DESKiDS2023}. In the DES Y3 cosmic shear analysis, which uses additional calibration data to break the IA--redshift uncertainty degeneracy, the NLA and TATT posteriors differ by
$0.5\sigma$ in $S_8$, and extending the TATT analysis to smaller scales removes this difference
\cite{Secco2022}.
 
These degeneracies motivate a profile-likelihood analysis of cosmic shear. So far, related ideas have been used mainly to marginalize analytically over nuisance parameters for speed~\cite{Hadzhiyska2023}. Public profile likelihood codes such as \textsc{pinc}~\cite{Herold2025}, \textsc{Procoli}~\cite{Karwal2024} and \textsc{Prospect}~\cite{Holm2024PROSPECT} rely on simulated annealing, while combining neural-network emulators with gradient-based optimizers reduces the cost further~\cite{Nygaard2023}. A full profile requires a global optimization at each value of the profiled parameter, on a likelihood with strong degeneracies, extended flat directions and multiple local minima. For cosmic shear, each optimization needs many evaluations of the nonlinear matter power spectrum, which dominate the cost. 
 
In this work, we build a computationally efficient profile likelihood pipeline that couples CMA-ES optimization to a fast cosmic shear likelihood evaluation in \textsc{CoCoA}~\cite{cocoa} using a matter power spectrum emulator. We validate the analysis on noiseless and noisy synthetic data. We then apply it to DES Y3 cosmic shear with the KiDS-overlapping area excised and the Hybrid model of Ref.~\cite{DESKiDS2023}, which allows a direct comparison with the published Bayesian results. We then test the robustness of the profile constraints to the range of $T_\mathrm{AGN}$ and the inclusion of small scales, to the IA model (\NLA{} or TATT, with $a_1$ free or restricted to positive values) and the addition of the DES Y3 shear ratios, which constrain the source redshifts and IA independently of cosmic shear~\cite{y3-shearratio}, to the sum of neutrino masses, and to the addition of CMB likelihood.
 
\section{Parameter inference: Bayesian vs frequentist}
\label{sec:stats}
\begin{figure*}[t]
    \centering
    \includegraphics[width=1.0\linewidth]{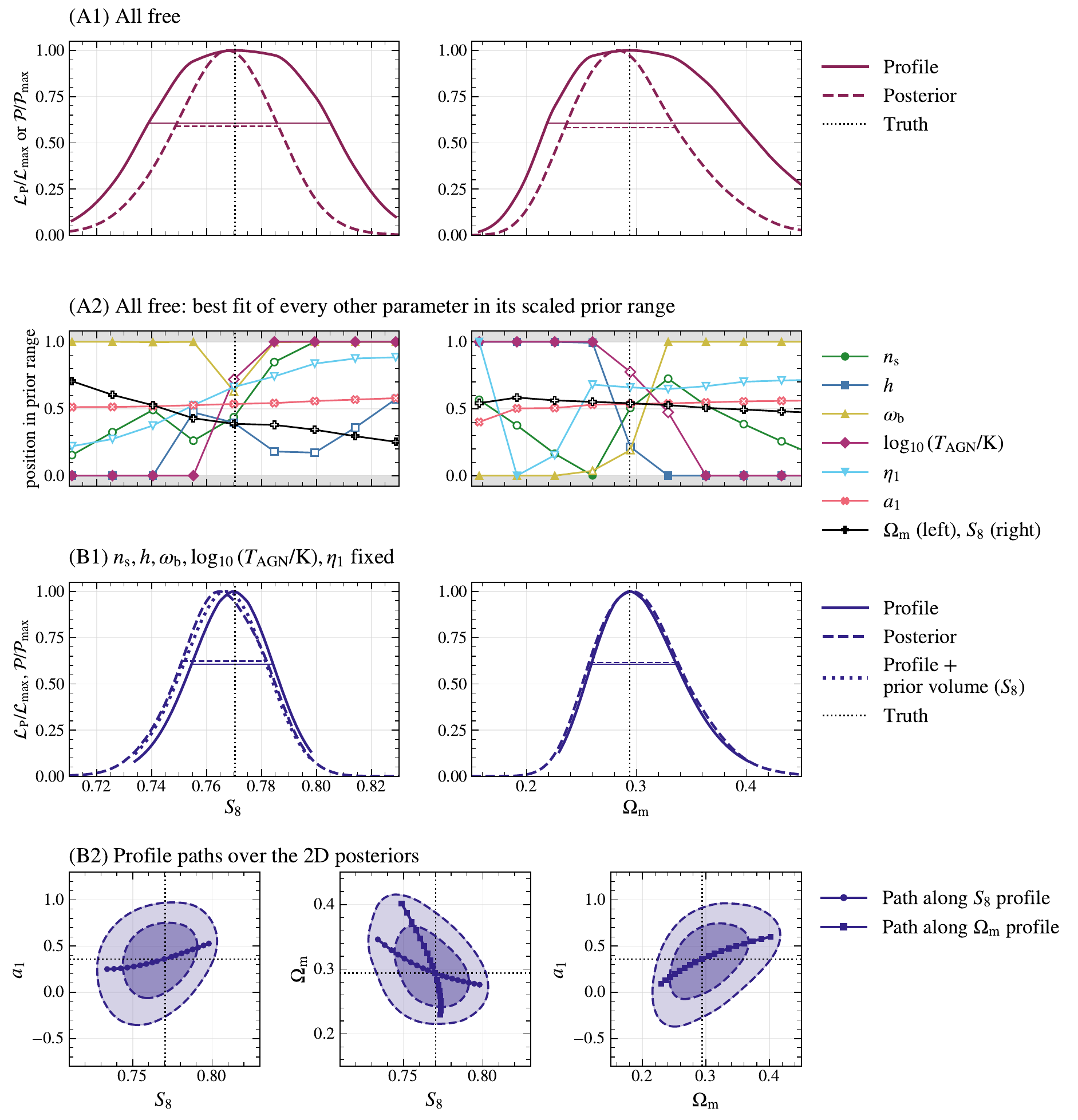}
    \caption{Comparison of Bayesian and frequentist approaches on one noiseless synthetic DES~Y3-like cosmic shear
data vector with $\Lambda$CDM-optimized scale cuts, generated with NLA. Thin black
dotted lines mark the input values used to generate the data vector.
\emph{First row:} With all parameters free, the profile likelihoods yield weaker constraints than the marginal posteriors. Horizontal bars, in the line style of
their curve, show the $1\sigma$ intervals of the profile ($\Delta\chi^2 = 1$) and the posterior.
\emph{Second row:} Best-fit values of all other free parameters along the $S_8$ (left) and
$\Omega_\mathrm{m}$ (right) profiles of the first row, shown as their position within the parameter range
(0 and 1 are the prior edges; grey bands lie outside the prior). Filled symbols mark best fits for unconstrained parameters along the profile that are located at the edges of their parameter ranges. The negligible prior volume near the edges suppresses the marginal posteriors relative to the profile likelihoods in the first row. 
\emph{Third row:} With parameters other than $S_8$, $\Omega_\mathrm{m}$ and $a_1$ fixed, the profiles and posteriors agree closely. The small offset between the $S_8$ profile and posterior is a prior-volume effect: once the Laplace estimate of the
prior-volume term is included (dotted), the volume-corrected profile and the posterior agree closely.
\emph{Fourth row:} Profile trajectories of the fixed-parameter analysis over 2D posteriors.
Circles follow the $S_8$ profile and squares the $\Omega_\mathrm{m}$ profile; both are shown in the
middle panel. The trajectories follow the degeneracy direction of the posterior. The posterior volume is
larger at low $S_8$, which explains the slight shift in the $S_8$ posterior relative to the profile in the third row.}
\label{fig:noiseless_data_profile}
\end{figure*}

\subsection{Bayesian inference}

The quantity of interest is the posterior distribution of the model
parameters $\boldsymbol{\theta} = (\boldsymbol{\Theta},\boldsymbol{n})$,
where $\boldsymbol{\Theta}$ are the parameters of interest and
$\boldsymbol{n}$ denotes the nuisance parameters. It is given by Bayes'
theorem,
\begin{equation}
    \mathcal{P}(\boldsymbol{\theta}\,|\,\boldsymbol{d}) \propto \mathcal{L}(\boldsymbol{d}\,|\,\boldsymbol{\theta})\,\Pi(\boldsymbol{\theta})\,,
\end{equation} 
 where $\mathcal{L}(\boldsymbol{d}\,|\,\boldsymbol{\theta})$ is the
likelihood of the data vector $\boldsymbol{d}$ and $\Pi(\boldsymbol{\theta})$ is the prior. Constraints on $\boldsymbol{\Theta}$ are obtained using the marginal posterior by integrating the full posterior over $\boldsymbol{n}$,
\begin{equation}
    \mathcal{P}(\boldsymbol{\Theta}\,|\,\boldsymbol{d}) \propto
    \int \mathcal{L}(\boldsymbol{d}\,|\,\boldsymbol{\Theta},\boldsymbol{n})\,
    \Pi(\boldsymbol{\Theta},\boldsymbol{n})\,\mathrm{d}\boldsymbol{n}\,.
\end{equation}
The posterior assigns a probability directly to 
$\boldsymbol{\Theta}$. Therefore, the 1D intervals are quoted as regions of fixed posterior mass: the $1\sigma$
interval on a parameter is the range containing $68.3\%$ of the marginal
posterior, and the $2\sigma$ interval the range containing $95.4\%$.

\subsection{Frequentist inference}
The frequentist approach maximizes the likelihood as a function of the parameter of interest without introducing any priors. The profile likelihood for $\boldsymbol{\Theta}$ is obtained by maximizing the likelihood over $\boldsymbol{n}$,
\begin{equation}
    \mathcal{L}_{\rm p}(\boldsymbol{\Theta}) \equiv
    \max_{\boldsymbol{n}} \mathcal{L}(\boldsymbol{d}\,|\,\boldsymbol{\Theta},\boldsymbol{n})
    = \mathcal{L}\big(\boldsymbol{d}\,|\,\boldsymbol{\Theta},
    \hat{\boldsymbol{n}}(\boldsymbol{\Theta})\big)\,,
\end{equation}
where $\hat{\boldsymbol{n}}(\boldsymbol{\Theta})$ denotes the value of
$\boldsymbol{n}$ that maximizes the likelihood at fixed $\boldsymbol{\Theta}$.
The confidence intervals are constructed from the likelihood ratio test
statistic,
\begin{equation}
    \Delta\chi^{2}(\boldsymbol{\Theta}) \equiv
    -2\ln\left[\frac{\mathcal{L}_{\rm p}(\boldsymbol{\Theta})}
    {\mathcal{L}_\mathrm{max}}\right]
    = \chi^{2}\big(\boldsymbol{\Theta},\hat{\boldsymbol{n}}(\boldsymbol{\Theta})\big)
    - \chi^{2}_{\rm min}\,,
\end{equation}
where $\mathcal{L}_\mathrm{max} \equiv \mathcal{L}(\boldsymbol{d}\,|\,\hat{\boldsymbol{\theta}})$ denotes the global maximum likelihood point at $\hat{\boldsymbol{\theta}}$. For a flat prior on $\boldsymbol{n}$, the marginal posterior and the profile likelihood are related by
\begin{equation}
    \mathcal{P}(\boldsymbol{\Theta}) \propto \mathcal{L}_\mathrm{p}(\boldsymbol{\Theta})\,V(\boldsymbol{\Theta})\,,
    \qquad
    V(\boldsymbol{\Theta}) \equiv \int \frac{\mathcal{L}(\boldsymbol{d}\,|\,\boldsymbol{\Theta},\boldsymbol{n})}
    {\mathcal{L}_\mathrm{p}(\boldsymbol{\Theta})}\,\mathrm{d}\boldsymbol{n}\,,
    \label{eq:volume}
\end{equation}
where $V(\boldsymbol{\Theta})$ is the volume of nuisance-parameter space that fits the data well at fixed $\boldsymbol{\Theta}$. The posterior is the profile likelihood weighted by this volume. We illustrate the different scenarios comparing the Bayesian and Frequentist approaches in a toy model in App.~\ref{app:toy}.

Frequentist confidence intervals are  set by a threshold on the test statistic $\Delta\chi^2$. For a single parameter of interest, the $1\sigma$  interval is the range with $\Delta\chi^{2} \leq 1$, and the $2\sigma$ interval is the range with $\Delta\chi^{2} \leq 4$. Joint two-dimensional regions are obtained by profiling over all remaining parameters on a grid in the two parameters of interest, and the corresponding thresholds are $\Delta\chi^{2} \leq 2.30$ and $\Delta\chi^{2} \leq 6.18$, since the test statistic then follows a $\chi^2$ distribution with two degrees of freedom. In both cases, the number of degrees of freedom is set by the number of parameters of interest. This mapping relies on Wilks' theorem~\cite{Wilks1938}, whose asymptotic assumptions can break down when the likelihood is strongly non-Gaussian, when parameters are poorly constrained by the data, or when the best fit lies at the edge of the allowed parameter range. In such scenarios, the thresholds must be calibrated explicitly, with a Neyman construction based on mock data sets~\cite{Neyman1937} or, near a physical boundary, the Feldman--Cousins prescription~\cite{FeldmanCousins1998}, to guarantee correct frequentist coverage as shown in~\cite{Herold2025} in the context of CMB data.

\subsection{Numerical implementation}
For cosmic shear, the likelihood minimization is difficult because the full parameter space has 16 (NLA) or 19 (TATT) dimensions, several unconstrained parameters sit at the edges of their parameter ranges at the optimum, and the likelihood has multiple local minima. We use the covariance matrix adaptation evolution strategy (CMA-ES)~\cite{Hansen2001, Hansen2016, pycma}. At each step, it draws a population of trial points from a multivariate Gaussian, ranks them by $\chi^2$, and moves the mean towards the best points while updating the step size and the full covariance of the Gaussian from the successful steps. The covariance therefore aligns with the degeneracy directions of the likelihood, approaching the global minimum efficiently. The CMA-ES algorithm is about nine times faster than the simulated-annealing code \textsc{Procoli}~\cite{Karwal2024} (Fig.~\ref{fig:emulator_camb} in App.~\ref{app:algorithm}).

\subsection{Comparison on synthetic data}
We compare the Bayesian and frequentist approaches on a noiseless synthetic cosmic shear data vector generated for a DES~Y3-like survey with $\Lambda$CDM scale cuts in Fig.~\ref{fig:noiseless_data_profile}. With all parameters free as shown in Panel~(A1), the profile likelihood is substantially broader than the marginal posterior and the two peaks do not coincide exactly. The reason is provided in Panel~(A2) where several parameters not constrained by cosmic shear are driven to the edge of the chosen parameter range. Since the prior volume is negligible at the edges, the posteriors shrink compared to the profiles. Once the weakly constrained parameters are fixed in Panel~(B1), the profile likelihood and marginal posteriors are in good agreement.  Panel~(B2) shows that the best-fit values along the profile track the same degeneracy direction as the posterior.

We validate the pipeline on synthetic data in App.~\ref{app:validation}. We test five input cosmologies drawn from a Latin hypercube (Fig.~\ref{fig:noiseless_cosmo}) and ten noisy realizations of the synthetic data vector (Fig.~\ref{fig:noisy-data-profile}). For both IA models, the profile best fit recovers the input cosmology, and the interval obtained by averaging over the ten noise realizations contains the true values of $S_8$ and $\Omega_\mathrm{m}$. Note that profile likelihoods are not generally expected to be broader than posteriors. In the case of cosmic shear, this broadening occurs because the best-fit parameters lie near the boundaries of the optimization parameter space, introducing edge effects into the profile construction (see App.~\ref{app:toy} for details). 

\section{Data and theoretical modeling}
\label{sec:model}

We adopt the DES Y3 + KiDS-1000 `Hybrid' model of Ref.~\cite{DESKiDS2023} as our baseline and therefore analyze the KiDS-excised DES Y3 cosmic shear data vector and covariance of that work. This allows us to compare our profile likelihoods directly with the published Hybrid posteriors. We describe the data in Sec.~\ref{sec:data}, and the baseline model, including the \NLA{} IA model, in Secs.~\ref{sec:model_2pt}--\ref{sec:model_lkl}. The TATT IA model, which we use only as a robustness test, is described in Sec.~\ref{sec:model_tatt}. We use the notation of~\cite{DESKiDS2023} throughout, supplemented by~\cite{Secco2022,Krause2021} for additional IA modeling details.

\subsection{KiDS-excised DES Y3 cosmic shear data}
\label{sec:data}

To reduce the cross-covariance between the two surveys, Ref.~\cite{DESKiDS2023} removes the region that overlaps with the KiDS-1000 footprint from the DES Y3 footprint. This removes $337\,\mathrm{deg}^2$ of unmasked area, or $8\%$ of the DES Y3 data. The redshift distributions of the four tomographic source bins are assumed to be unchanged. The effective number densities, shape noise, and shear responses are re-measured on the reduced footprint, and the analytic covariance~\cite{Krause2017,Fang2020, y3-covariances} is recomputed~\cite{DESKiDS2023}. In the Hybrid analysis, removing the overlap region changes $S_8$ by $0.004$~\cite{DESKiDS2023}. The data vector consists of the auto- and cross-correlation functions $\xi^{(ij)}_\pm(\theta)$ of the four source bins in 20 logarithmic angular bins between $2.5'$ and $250'$. As in Ref.~\cite{DESKiDS2023}, the baseline analysis uses the DES Y3 `$\Lambda$CDM-optimized' scale cuts~\cite{Amon2022,Secco2022}, which retain 273 data points (184 in $\xi_+$ and 89 in $\xi_-$)~\cite{Secco2022}. In Sec.~\ref{sec:baryons} we also include all scales $\theta \geq 2.5'$.

\subsection{Modeling cosmic shear}
\label{sec:model_2pt}

The shear two-point correlation functions are related to the E- and B-mode cosmic shear power spectra, $C^\mathrm{EE}_{\epsilon\epsilon}$ and $C^\mathrm{BB}_{\epsilon\epsilon}$, by~\cite{DESKiDS2023}
\begin{align}
    \xi^{(ij)}_\pm(\theta) = \frac{1}{4\pi}\sum_{\ell\geq2}(2\ell+1)\,&d^\ell_{2,\pm2}(\theta) \nonumber\\
    &\times\left[C^{\mathrm{EE},(ij)}_{\epsilon\epsilon}(\ell) \pm C^{\mathrm{BB},(ij)}_{\epsilon\epsilon}(\ell)\right],
    \label{eq:xipm}
\end{align}
where $d^\ell_{m,n}$ are the reduced Wigner $d$-matrices. We average the model over each angular bin~\cite{Krause2021}. Including intrinsic alignments (IA), the shear power spectra of tomographic bins $i$ and $j$ are~\cite{Secco2022}
\begin{align}
    C^{\mathrm{EE},(ij)}_{\epsilon\epsilon}(\ell) &= C^{(ij)}_\mathrm{GG}(\ell) + C^{(ij)}_\mathrm{GI}(\ell) + C^{(ji)}_\mathrm{GI}(\ell) + C^{(ij)}_\mathrm{II,EE}(\ell)\,, \nonumber\\
    C^{\mathrm{BB},(ij)}_{\epsilon\epsilon}(\ell) &= C^{(ij)}_\mathrm{II,BB}(\ell)\,,
    \label{eq:Cl_EE_BB}
\end{align}
where G denotes gravitational shear and I intrinsic shape. 
In the flat-sky Limber approximation, the lensing term is~\cite{DESKiDS2023}
\begin{equation}
    C^{(ij)}_\mathrm{GG}(\ell) = \int_0^{\chi_\mathrm{hor}} \mathrm{d}\chi\, \frac{W^{(i)}(\chi)\,W^{(j)}(\chi)}{f_\mathrm{K}^2(\chi)}\, P_\delta\!\left(\frac{\ell+1/2}{f_\mathrm{K}(\chi)},\, \chi\right),
    \label{eq:limber}
\end{equation}
with $P_\delta$ the nonlinear matter power spectrum, $\chi$ the comoving distance, $\chi_\mathrm{hor}$ the comoving distance to the horizon, and $f_\mathrm{K}(\chi)$ the comoving angular diameter distance, which equals $\chi$ in the flat models considered here. The lensing efficiency of source bin $i$ is
\begin{equation}
    W^{(i)}(\chi) = \frac{3\,\Omega_\mathrm{m} H_0^2}{2\,c^2}\,\frac{\chi}{a(\chi)} \int_\chi^{\chi_\mathrm{hor}} \mathrm{d}\chi'\, n_i(\chi')\,\frac{\chi'-\chi}{\chi'}\,,
    \label{eq:lens_eff}
\end{equation}
where $c$ is the speed of light, $n_i(\chi) = n_i(z)\,\mathrm{d}z/\mathrm{d}\chi$ is the source redshift distribution, normalized to unity. The IA terms are~\cite{Secco2022}
\begin{align}
    C^{(ij)}_\mathrm{GI}(\ell) &= \int_0^{\chi_\mathrm{hor}} \mathrm{d}\chi\, \frac{W^{(i)}(\chi)\,n_j(\chi)}{\chi^2}\, P_\mathrm{GI}\!\left(\frac{\ell+1/2}{\chi},\, \chi\right), \nonumber\\
    C^{(ij)}_\mathrm{II}(\ell) &= \int_0^{\chi_\mathrm{hor}} \mathrm{d}\chi\, \frac{n_i(\chi)\,n_j(\chi)}{\chi^2}\, P_\mathrm{II}\!\left(\frac{\ell+1/2}{\chi},\, \chi\right),
    \label{eq:limber_ia}
\end{align}
where the second line applies to both the E- and B-mode II spectra, and $P_\mathrm{GI}$ and $P_\mathrm{II}$ are given by the IA model (Sec.~\ref{sec:model_ia}).

\subsection{Matter power spectrum and baryon feedback}
\label{sec:model_pk}

We compute $P_\delta$ with \textsc{HMCode2020}~\cite{Mead2021} from \textsc{CAMB}~\cite{Lewis2000} linear power spectra; \textsc{HMCode2020} includes massive neutrinos. Baryonic feedback, which suppresses the matter power spectrum on small scales, is described by one parameter, $T_\mathrm{AGN}$, the AGN heating temperature in the BAHAMAS hydrodynamical simulations~\cite{McCarthy2017} to which \textsc{HMCode2020} is calibrated. We sample $\log_{10}(T_\mathrm{AGN}/\mathrm{K})$. 
As in Ref.~\cite{DESKiDS2023}, the lower prior boundary, $\log_{10}(T_\mathrm{AGN}/\mathrm{K})=7.3$, gives cosmic shear predictions close to a gravity-only model, and the upper boundary, $8.0$, is the strongest feedback for which BAHAMAS matches the observed baryon fractions of galaxy groups. To speed up the profile likelihoods, we use the KiDS-Legacy \textsc{CosmoPower}~\cite{SpurioMancini2022, Wright2025} emulator of \textsc{HMCode2020} instead of \textsc{CAMB}+\textsc{HMCode2020}~\cite{Wright2025} and validate the emulator against \textsc{CAMB} in Figs.~\ref{fig:bmcmc_comparison} and~\ref{fig:emulator_camb}. We compute angular power spectra and correlation functions with \textsc{CosmoLike}~\cite{Krause2017}. We adopt the \textsc{CoCoA}~\cite{cocoa} inference pipeline for likelihood evaluations and Bayesian inference.

\subsection{Intrinsic alignments: \NLA{} baseline}
\label{sec:model_ia}

As in Ref.~\cite{DESKiDS2023}, our baseline IA model is the nonlinear linear alignment model with power-law redshift evolution (\NLA{})~\cite{Hirata2004,Bridle2007}. All results use \NLA{} unless stated otherwise. In \NLA{}, intrinsic shapes are linear in the tidal field, and the IA power spectra are proportional to the nonlinear matter power spectrum,
\begin{align}
    P_\mathrm{GI}(k,z) &= A_1(z)\,P_\delta(k,z)\,, \nonumber\\
    P_\mathrm{II,EE}(k,z) &= A_1^2(z)\,P_\delta(k,z)\,,
    \label{eq:nla_pk}
\end{align}
with $P_\mathrm{II,BB}=0$. The amplitude is~\cite{DESKiDS2023,Secco2022}
\begin{equation}
    A_1(z) = -a_1\,\bar{C}_1\,\frac{\rho_\mathrm{crit}\Omega_\mathrm{m}}{D(z)}\left(\frac{1+z}{1+z_\mathrm{pivot}}\right)^{\eta_1},
    \label{eq:nla_amp}
\end{equation}
where $D(z)$ is the linear growth factor, $\rho_\mathrm{crit}$ the critical density, $\bar{C}_1 = 5\times10^{-14}\,M_\odot^{-1} h^{-2}\,\mathrm{Mpc}^3$ a normalization constant fixed by convention, and $z_\mathrm{pivot} = 0.62$~\cite{DESKiDS2023}. The free parameters are the amplitude $a_1$ and the power-law index $\eta_1$, denoted $A_\mathrm{IA}$ and $\eta_\mathrm{IA}$ in Ref.~\cite{DESKiDS2023}; we use $a_1$ and $\eta_1$ so that the notation is the same as for TATT (Sec.~\ref{sec:model_tatt}). With this sign convention, $a_1>0$ means that galaxies are aligned with the stretching direction of the tidal field, as expected for tidal alignment. We restrict our baseline analysis to $a_1\geq0$, based on hydrodynamical simulations, direct
observations and the shear ratios (Sec.~\ref{sec:analysis_a1}).

\subsubsection{TATT extension}
\label{sec:model_tatt}

We use the `tidal alignment and tidal torquing' (TATT) model~\cite{Blazek2019}, the baseline IA model of the DES Y3 analyses~\cite{Krause2021,Secco2022}, only to test how our results depend on the IA model (Sec.~\ref{sec:ia_robustness}). TATT expands the intrinsic shape field in the density and tidal field,
\begin{equation}
    \tilde{\gamma}_{\alpha,\mathrm{IA}} = A_1 s_\alpha + A_{1\delta}\,\delta\, s_\alpha + A_2\,(s\times s)_\alpha + \cdots\,.
\end{equation}
The first term is the NLA term, the second term accounts for the weighting of the shape field by the source density, and the third term, quadratic in the tidal field, describes tidal torquing. Suppressing the redshift dependence of the coefficients, the IA power spectra are~\cite{Blazek2019,Secco2022}
\begin{align}
    P_\mathrm{GI} &= A_1 P_\delta + A_{1\delta} P_{0|0\mathrm{E}} + A_2 P_{0|\mathrm{E}2}\,, \nonumber\\
    P_\mathrm{II,EE} &= A_1^2 P_\delta + 2A_1A_{1\delta} P_{0|0\mathrm{E}} + A_{1\delta}^2 P_{0\mathrm{E}|0\mathrm{E}} + A_2^2 P_{\mathrm{E}2|\mathrm{E}2} \nonumber\\
    &\quad + 2A_1A_2 P_{0|\mathrm{E}2} + 2A_{1\delta}A_2 P_{0\mathrm{E}|\mathrm{E}2}\,, \nonumber\\
    P_\mathrm{II,BB} &= A_{1\delta}^2 P_{0\mathrm{B}|0\mathrm{B}} + A_2^2 P_{\mathrm{B}2|\mathrm{B}2} + 2A_{1\delta}A_2 P_{0\mathrm{B}|\mathrm{B}2}\,,
    \label{eq:tatt_pk}
\end{align}
where the $k$-dependent terms are one-loop integrals over the linear matter power spectrum, which we compute with \textsc{FAST-PT}~\cite{McEwen2016,Fang2017}. Unlike \NLA{}, TATT produces B-mode power. The amplitude $A_1$ is given by Eq.~\eqref{eq:nla_amp}, and~\cite{Secco2022}
\begin{align}
    A_2(z) &= 5a_2\,\bar{C}_1\,\frac{\rho_\mathrm{crit}\Omega_\mathrm{m}}{D^2(z)}\left(\frac{1+z}{1+z_\mathrm{pivot}}\right)^{\eta_2}, \nonumber\\
    A_{1\delta}(z) &= b_\mathrm{TA}\,A_1(z)\,,
    \label{eq:tatt_amp}
\end{align}
where $b_\mathrm{TA}$ is the linear bias of the source galaxies, taken to be constant in redshift. TATT has five free parameters: the tidal alignment amplitude $a_1$ and its redshift evolution $\eta_1$, the tidal torquing amplitude $a_2$ and its redshift evolution $\eta_2$, and $b_\mathrm{TA}$. We use the DES Y3 priors for these (Table~\ref{tab:params}).

\NLA{} and TATT are nested models: for $a_2\to0$ and $b_\mathrm{TA}\to0$, TATT reduces to \NLA{}, and $\eta_2$ has no effect~\cite{DESKiDS2023,Secco2022}. Both limits are needed: for fixed non-zero $b_\mathrm{TA}$, TATT does not reduce to \NLA{} as $a_2\to0$, because the density-weighting term $A_{1\delta}=b_\mathrm{TA}A_1$ still adds the $P_{0|0\mathrm{E}}$, $P_{0\mathrm{E}|0\mathrm{E}}$, and $P_{0\mathrm{B}|0\mathrm{B}}$ contributions in Eq.~\eqref{eq:tatt_pk}, including B-mode power.

\subsection{Observational systematics}
\label{sec:model_sys}

As in Ref.~\cite{DESKiDS2023}, we use the DES Y3 treatment of calibration uncertainties:
For each source bin, the uncertainty in the redshift calibration is modeled with a shift parameter $\Delta z_i$, such that
$n_i(z) \rightarrow n_i(z - \Delta z_i)$, i.e.\ $\Delta z_i = \langle z\rangle^{\rm true}_i - \langle z\rangle^{\rm estimate}_i$~\cite{Myles2021,Amon2022}\footnote{We follow the DES Y3 cosmic shear convention $n_i(z)\rightarrow n_i(z-\Delta z_i)$~\cite{Amon2022,Secco2022}, so $\Delta z_i>0$ moves bin $i$ to higher redshift. Ref.~\cite{DESKiDS2023} defines $\Delta z_i$ with the opposite sign, as $\langle n_i^\mathrm{estimate}\rangle - \langle n_i^\mathrm{true}\rangle$. The priors are centred on zero and symmetric, so this only flips the sign of reported best-fit values.}, and 
the multiplicative shear calibration uncertainty is modeled as $\xi^{(ij)}_\pm \rightarrow (1+m_i)(1+m_j)\,\xi^{(ij)}_\pm$~\cite{MacCrann2022,Amon2022}.
Both $\Delta z_i$ and $m_i$ are calibrated with data and simulations external to the cosmic shear two-point analysis: the shear calibration with image simulations~\cite{MacCrann2022}, and the redshift calibration with self-organizing-map photometric redshifts combined with clustering redshifts~\cite{Myles2021,y3-sourcewz}. The Hybrid model~\cite{DESKiDS2023} does not include the DES Y3 shear-ratio likelihood~\cite{y3-shearratio}, which in the DES Y3 analyses additionally constrains the redshift and IA parameters~\cite{Amon2022,Secco2022}. We follow the Hybrid model for our baseline analysis, but add the shear-ratio likelihood to analysis variants that profile IA parameters (Sec.~\ref{sec:model_SR}). Additive shear biases are removed at the catalog level by subtracting the mean ellipticity in each tomographic bin~\cite{y3-shapecatalog}.

\subsection{Likelihood and parameters}
\label{sec:model_lkl}

We assume a Gaussian likelihood,
\begin{equation}
    -2\ln\mathcal{L}(\boldsymbol{d}\,|\,\boldsymbol{\theta}) = \left[\boldsymbol{d}-\boldsymbol{M}(\boldsymbol{\theta})\right]^\mathrm{T}\mathsf{C}^{-1}\left[\boldsymbol{d}-\boldsymbol{M}(\boldsymbol{\theta})\right] + \mathrm{const.}\,,
    \label{eq:lkl}
\end{equation}
where $\boldsymbol{M}(\boldsymbol{\theta})$ is the model data vector of Eq.~\eqref{eq:xipm} including systematics, and $\mathsf{C}$ is the analytic covariance of the KiDS-excised data vector~\cite{DESKiDS2023}. As in Ref.~\cite{DESKiDS2023}, we sample the cosmological parameters $\{S_8,\,\omega_\mathrm{c},\,\omega_\mathrm{b},\,h,\,n_\mathrm{s}\}$, where $S_8\equiv\sigma_8\sqrt{\Omega_\mathrm{m}/0.3}$, $\omega_\mathrm{c}\equiv\Omega_\mathrm{c}h^2$, $\omega_\mathrm{b}\equiv\Omega_\mathrm{b}h^2$, and $h$ is the dimensionless Hubble parameter. The parameters $\Omega_\mathrm{m}$ and $\sigma_8$ are derived. While~\cite{DESKiDS2023} also samples the sum of neutrino masses, $\Sigma m_\nu$, in our baseline analyses we instead fix $\Sigma m_\nu = 0.06\,\mathrm{eV}$, the value at which the power spectrum emulator was trained, as this enables an order-of-magnitude computational speed-up (cf.\ Fig.~\ref{fig:emulator_camb}). In App.~\ref{app:mnu}, we vary $\Sigma m_\nu$ at the cost of running CAMB instead of the emulator for the matter power spectrum. 
All other priors are those of Ref.~\cite{DESKiDS2023} and are listed in Table~\ref{tab:params}. For the profile likelihoods, the uniform priors set the parameter boundaries of the maximization. The redshift and shear calibration parameters are informed by external calibration data, which in a Bayesian analysis enter as Gaussian priors. A profile likelihood has no priors, and the consistent frequentist approach would be to fit the calibration data jointly with the shear data, but this is computationally not feasible. Instead, we approximate each calibration result as a Gaussian measurement of the corresponding nuisance parameter, with mean and width from Table~\ref{tab:params}, and add it to the likelihood,
\begin{equation}
    \chi^2_\mathrm{nuisance} = \sum_{i=1}^{4}\left(\frac{\Delta z_i-\mu_{\Delta z,i}}{\sigma_{\Delta z,i}}\right)^2
    + \sum_{i=1}^{4}\left(\frac{m_i-\mu_{m,i}}{\sigma_{m,i}}\right)^2 .
    \label{eq:chi2_nuisance}
\end{equation}
This is the standard treatment of auxiliary measurements in frequentist analyses~\cite{Cowan2011,Cranmer2015}. Since each Gaussian term adds one measurement and one parameter, it introduces no prior-volume effect and does not change the number of degrees of freedom of $\Delta\chi^2$ (Sec.~\ref{sec:stats}). Profile analyses of galaxy clustering have treated external constraints the same way, for EFT nuisance parameters~\cite{Holm2023} and for BBN and $n_\mathrm{s}$~\cite{Morawetz2025}. The approximation assumes that the calibration results are dominated by the calibration data rather than by the priors of those analyses.

When combining cosmic shear with the CMB, we add a compressed CMB Gaussian likelihood for $(\omega_\mathrm{b},\, n_\mathrm{s},\, h)$ derived with mean and covariance taken from the \textit{Planck} 2018 high-$\ell$ TT,TE,EE chains~\cite{Planck2018}. Cosmic shear alone is not sensitive to these parameters; however, they set the shape of the linear matter power spectrum, so constraining them with CMB breaks their degeneracies with $\Omega_\mathrm{m}$ and $S_8$ and increases the constraining power of the analysis.
\begin{table}[t]
\centering
\caption{Model parameters and priors. $\mathcal{U}[a,b]$ denotes a uniform prior and $\mathcal{N}(\mu,\sigma)$ a Gaussian prior with mean $\mu$ and standard deviation $\sigma$. All priors are those of the DES Y3 + KiDS-1000 Hybrid analysis~\cite{DESKiDS2023}, except for the fixed neutrino mass. The TATT and calibration priors are those of DES Y3~\cite{Secco2022}, which Ref.~\cite{DESKiDS2023} also uses. In the profile likelihoods, the Gaussian calibration priors enter as likelihood terms (Eq.~\ref{eq:chi2_nuisance}).}
\label{tab:params}
{\small
\setlength{\tabcolsep}{4pt}
\renewcommand{\arraystretch}{0.95}
\begin{tabular}{l l}
\toprule
Parameter & Prior \\
\midrule
\multicolumn{2}{l}{\textit{Cosmology}} \\
$S_8$ & $\mathcal{U}[0.1,\,1.3]$ \\
$\omega_\mathrm{c}$ & $\mathcal{U}[0.051,\,0.255]$ \\
$\omega_\mathrm{b}$ & $\mathcal{U}[0.019,\,0.026]$ \\
$h$ & $\mathcal{U}[0.64,\,0.82]$ \\
$n_\mathrm{s}$ & $\mathcal{U}[0.84,\,1.1]$ \\
$\Sigma m_\nu$ [eV] & $0.06$ (fixed)$^{a}$ \\
$\Omega_\mathrm{m}$, $\sigma_8$ & derived \\
\midrule
\multicolumn{2}{l}{\textit{Baryon feedback}} \\
$\log_{10}(T_\mathrm{AGN}/\mathrm{K})$ & $\mathcal{U}[7.3,\,8.0]$$^{b}$ \\
\midrule
\multicolumn{2}{l}{\textit{Intrinsic alignment (\NLA{}, baseline)}} \\
$a_1$ & $\mathcal{U}[-5,\,5]$ or $\mathcal{U}[0,\,5]$$^{c}$ \\
$\eta_1$ & $\mathcal{U}[-5,\,5]$ \\
\multicolumn{2}{l}{\textit{Intrinsic alignment (TATT extension, in addition)}} \\
$a_2$, $\eta_2$ & $\mathcal{U}[-5,\,5]$ \\
$b_\mathrm{TA}$ & $\mathcal{U}[0,\,2]$ \\
\midrule
\multicolumn{2}{l}{\textit{Redshift calibration}} \\
$\Delta z_1$ & $\mathcal{N}(0,\,0.018)$ \\
$\Delta z_2$ & $\mathcal{N}(0,\,0.015)$ \\
$\Delta z_3$ & $\mathcal{N}(0,\,0.011)$ \\
$\Delta z_4$ & $\mathcal{N}(0,\,0.017)$ \\
\midrule
\multicolumn{2}{l}{\textit{Shear calibration}} \\
$m_1$ & $\mathcal{N}(-0.0063,\,0.0091)$ \\
$m_2$ & $\mathcal{N}(-0.0198,\,0.0078)$ \\
$m_3$ & $\mathcal{N}(-0.0241,\,0.0076)$ \\
$m_4$ & $\mathcal{N}(-0.0369,\,0.0076)$ \\
\bottomrule
\end{tabular}
\begin{flushleft}
\footnotesize
$^{a}$ Hybrid prior $\mathcal{U}[0.055,\,0.6]$ used in App.~\ref{app:mnu}.\\
$^{b}$ Upper limit extended to $8.3$ in Sec.~\ref{sec:baryons}.\\
$^{c}$ The restricted prior is used for the `$a_1\geq0$' analyses.
\end{flushleft}
}
\end{table}

\subsubsection{Shear ratios}
\label{sec:model_SR}
For some analysis variants that profile the intrinsic-alignment parameters,
we go beyond the Hybrid setup and add the DES Y3 small-scale shear-ratio (SR) likelihood of
Ref.~\cite{y3-shearratio} for the MagLim lens sample, as in the DES Y3 cosmic shear
analysis~\cite{Amon2022, Secco2022}. The shear ratios constrain the source redshifts and intrinsic
alignments independently of cosmic shear and so reduce the degeneracies between photo-$z$ and IA
parameters~\cite{y3-shearratio}. The likelihood consists of nine ratios of galaxy--galaxy lensing
measurements around the same lens bin (lens bins 1--3) for different source bins, on scales of
2--6\,$h^{-1}$Mpc, with a Gaussian likelihood and fixed covariance. Since the ratios are independent
of the two-point measurements~\cite{y3-shearratio}, their $\chi^2_\mathrm{SR}$ is added to that of cosmic shear
and the nuisance parameters [Eq.~\eqref{eq:L_tot} below]. The ratios are modelled as in DES Y3~\cite{y3-shearratio, Krause2021},
with cosmology, intrinsic alignments, source redshifts and shear calibration shared with cosmic shear.
The lens galaxies in bin $i$ have a linear bias $b_i$, and lens magnification enters with fixed
coefficients $C^\mathrm{l}_i$~\cite{y3-2x2ptmagnification}. Uncertainties in the lens redshift
distributions are described by a shift $\Delta z^\mathrm{l}_i$ and a stretch $s_i$ about the mean redshift
$\langle z\rangle_i$, $n^\mathrm{l}_i(z) \rightarrow n^\mathrm{l}_i\big((z-\Delta z^\mathrm{l}_i-\langle z\rangle_i)/s_i
+ \langle z\rangle_i\big)/s_i$~\cite{y3-2x2ptaltlensresults}, with priors calibrated with clustering
redshifts~\cite{y3-lenswz}. These nine lens parameters enter only the SR likelihood; we internally minimize over them at each evaluation of $\chi^2_{\rm{SR}}$, including the external calibration likelihood for the lens photo-z parameters (analogous to Eq.~\ref{eq:chi2_nuisance}, using the priors in Table~\ref{tab:sr_params}). We evaluate the SR
likelihood with a fast re-implementation of the DES Y3 model. It agrees with the DES Y3 model to better than $1\%$ in $\chi^2_\mathrm{SR}$ for
$\chi^2_\mathrm{SR}<100$ for both NLA and TATT. The SR data vector and
covariance are measured on the full DES Y3 footprint; no KiDS-excised version exists, so the SR
likelihood does not exactly match the excised area of the cosmic shear data vector.

Collecting the terms above, the total $\chi^2$ minimized at each point of a profile is
\begin{equation}
    \chi^2_\mathrm{tot} = \chi^2_\mathrm{cosmic\,shear} + \chi^2_\mathrm{nuisance}
    + \chi^2_\mathrm{CMB} + \chi^2_\mathrm{SR}\,,
\label{eq:L_tot}
\end{equation}
where $\chi^2_\mathrm{cosmic\,shear} = -2\ln\mathcal{L}$ from Eq.~\eqref{eq:lkl} and
$\chi^2_\mathrm{nuisance}$ is given by Eq.~\eqref{eq:chi2_nuisance}. The last two terms are
included only where stated: $\chi^2_\mathrm{CMB}$ in the analyses labelled ``+CMB'', and
$\chi^2_\mathrm{SR}$ in those labelled ``+ shear ratios''.

\begin{table}[t]
\centering
\caption{Lens parameters of the shear-ratio likelihood for MagLim lens bins $i=1$--3, with the DES Y3
priors~\cite{y3-lenswz, y3-2x2ptmagnification}. All other parameters are shared with
cosmic shear (Table~\ref{tab:params}).}
\label{tab:sr_params}
{\small
\setlength{\tabcolsep}{4pt}
\renewcommand{\arraystretch}{0.95}
\begin{tabular}{l l}
\toprule
Parameter & Prior \\
\midrule
\multicolumn{2}{l}{\textit{Lens redshift shift}} \\
$\Delta z^\mathrm{l}_1$ & $\mathcal{N}(-0.009,\,0.007)$ \\
$\Delta z^\mathrm{l}_2$ & $\mathcal{N}(-0.035,\,0.011)$ \\
$\Delta z^\mathrm{l}_3$ & $\mathcal{N}(-0.005,\,0.006)$ \\
\midrule
\multicolumn{2}{l}{\textit{Lens redshift stretch}} \\
$s_1$ & $\mathcal{N}(0.975,\,0.062)$ \\
$s_2$ & $\mathcal{N}(1.306,\,0.093)$ \\
$s_3$ & $\mathcal{N}(0.870,\,0.054)$ \\
\midrule
\multicolumn{2}{l}{\textit{Linear galaxy bias}} \\
$b_i$ & $\mathcal{U}[0.8,\,3.0]$ \\
\midrule
\multicolumn{2}{l}{\textit{Lens magnification}} \\
$C^\mathrm{l}_1,\,C^\mathrm{l}_2,\,C^\mathrm{l}_3$ & $0.42,\,0.30,\,1.76$ (fixed) \\
\bottomrule
\end{tabular}
}
\end{table}

\section{Data analysis: DES Y3 (KiDS excised)}
\label{sec:analysis_results}
\subsection{Restriction of NLA parameter space}
\label{sec:analysis_a1}
Our main results use the $a_1\geq0$ branch of the NLA profiles. When $a_1$ is allowed to be negative, we find that the $a_1<0$ part of the $S_8$ profile ($S_8<0.78$ in the top panel of Fig.~\ref{fig:desy3_nla_tatt}, discussed in Sec.~\ref{sec:ia_robustness}) has a local minimum at $a_1\approx-2.7$ and $\eta_1\approx5$ (the prior edge). This implies a large negative alignment amplitude that grows quickly with redshift: the effective amplitude $a_1^\mathrm{eff}(z)\equiv a_1[(1+z)/(1+z_\mathrm{pivot})]^{\eta_1}$ reaches $-1.8$ at $z=0.5$ and $-4.6$ at $z=0.8$. In linear alignment theory, hydrodynamical simulations, and measurements of density--shape correlations~\cite{Catelan2001, Hirata2004, Chisari2015, Joachimi2011, Singh2015, Johnston2019,
Samuroff2023}, red galaxies are radially aligned ($a_1>0$). For blue galaxies, which dominate the source sample, the sign is less certain. Hydrodynamical simulations disagree on the sign for disc galaxies~\cite{Chisari2015, Chisari2025}, and
spiral galaxies in FLAMINGO are tangentially aligned, with $a_1=-0.76\pm0.16$ at
$z=0.5$~\cite{Kokron2026}. Measured alignment amplitudes of blue galaxies are consistent with zero,
with uncertainties of $\sigma(a_1)\simeq0.4$--1 and $|a_1|<0.78$ (95\% CL) for emission-line galaxies
at $z\approx0.8$~\cite{Mandelbaum2011, Johnston2019, Samuroff2023, Georgiou2025, Siegel2025}. A small
negative amplitude is possible, but not one as large as the $a_1^\mathrm{eff}$ found on the $a_1<0$ branch. In lensing data, the
alignment amplitude is degenerate with the amplitude of the lensing kernel and thus with the source
redshift calibration. The DES Y3 shear ratios reduce this degeneracy~\cite{y3-shearratio}. With them,
the best-fit $a_1$ is non-negative at every point of the NLA $S_8$ profile. In TATT, $a_1$ is degenerate
with the tidal-torquing amplitude $a_2$, and the best fit has a small negative $a_1$ and positive
$a_2$, as in DES Y3~\cite{Secco2022} (Sec.~\ref{sec:ia_robustness}). We also show profiles for the full range $a_1\in[-5,5]$ for comparison (Fig.~\ref{fig:desy3_nla_tatt} and Fig.~\ref{fig:desy3_lcdm_ia_free} in App.~\ref{app:ia}).

\subsection{Cosmic shear alone and combined with CMB}
\begin{figure}
    \centering
    \includegraphics[width=0.95\linewidth]{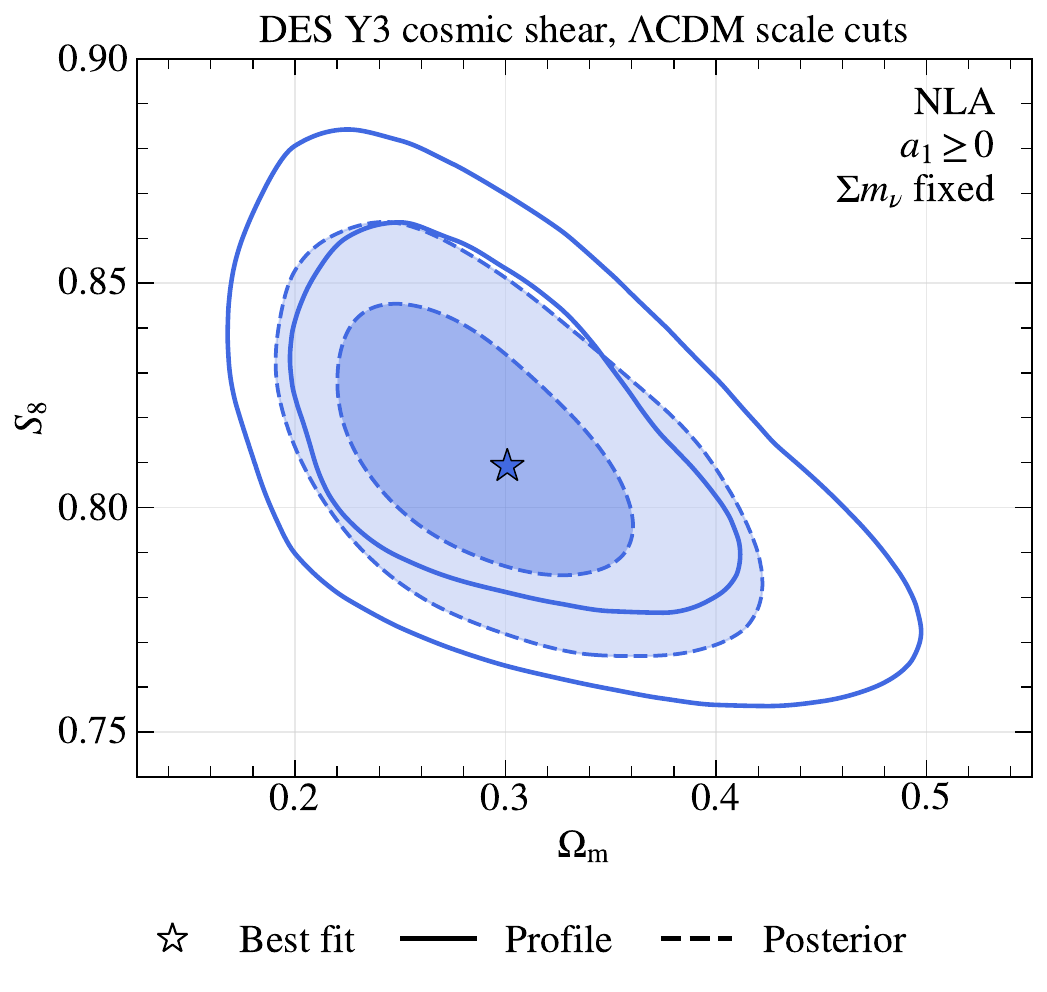}
    
    \caption{2D constraints from Bayesian (dashed contours) and profile likelihood (solid contours) analysis are consistent. Away from the best fit, the profile likelihood is maximized at the edges of the parameter ranges for poorly constrained parameters (see Fig.~\ref{fig:noiseless_data_profile} for intuition). These edges hold little prior volume, so they artificially shrink the marginal posteriors compared to the profiles.}
    \label{fig:desy3_s8_omegam}
\end{figure}
\begin{figure}
    \centering
    \includegraphics[width=1.0\linewidth]{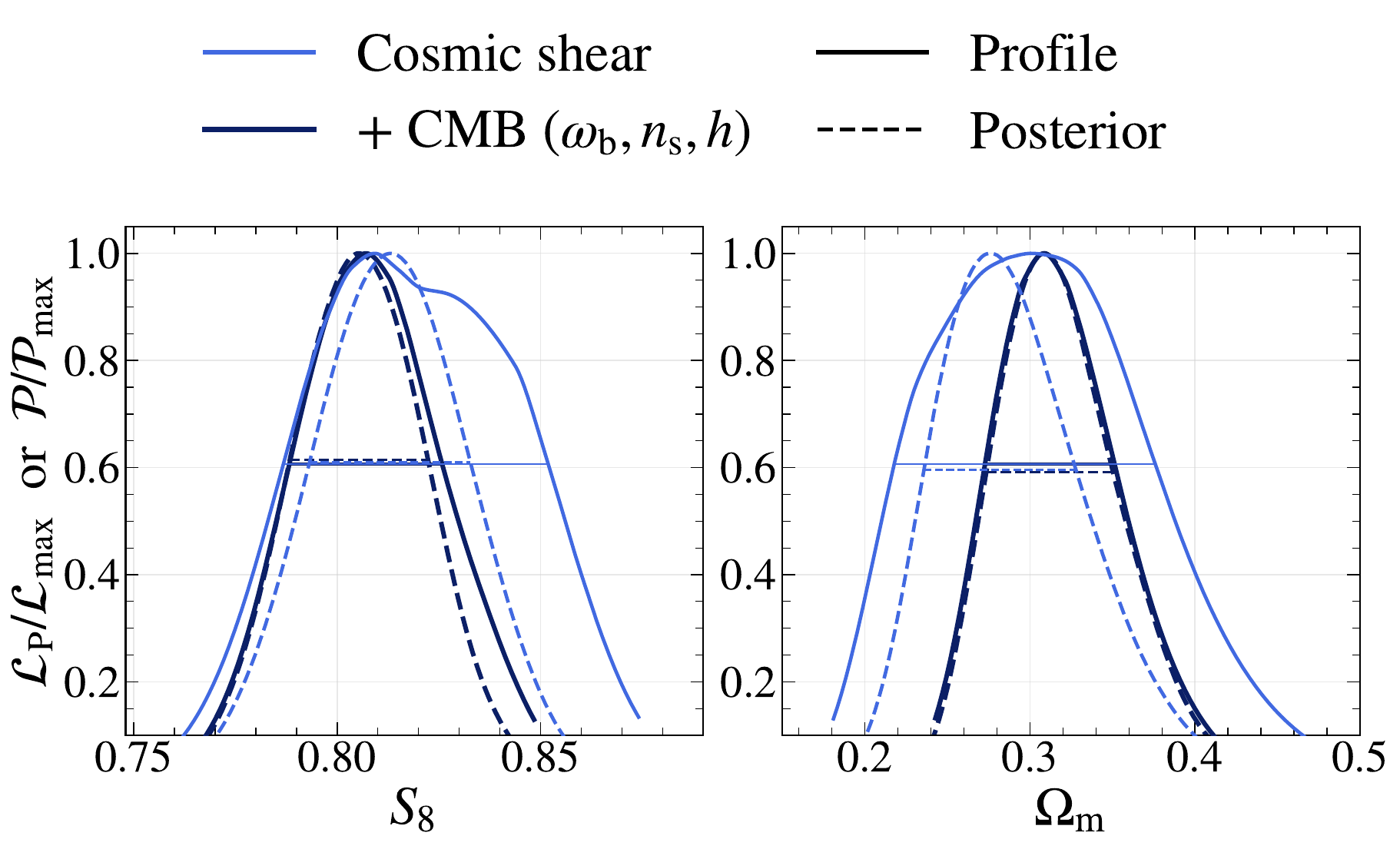}
    \caption{Addition of compressed CMB Planck 2018 TT,TE,EE Gaussian likelihood (in $\omega_\mathrm{b}$, $n_\mathrm{s}$, and $h$) to cosmic shear brings profiles and posteriors into agreement.}
    \label{fig:desy3_plus_cmb}
\end{figure}
Our main result from DES Y3 (KiDS excised) cosmic shear data, analyzed with the \NLA{} amplitude restricted to $a_1 \geq 0$ and $\Sigma m_\nu = 0.06$\,eV, is shown in Fig.~\ref{fig:desy3_s8_omegam}. Both the 2D profile likelihood constraints and the marginal posteriors enclose the maximum-likelihood point. We find the profile likelihood constraints to be weaker than the marginal posteriors because, away from the maximum-likelihood point, the likelihood is maximized at the edges of the parameter range. As described in Fig.~\ref{fig:noiseless_data_profile}, the prior volume is negligible at the boundaries, which makes the posteriors compact compared to the profiles.
 
Figure~\ref{fig:desy3_plus_cmb} shows the 1D constraints. For cosmic shear
alone, the $S_8$ and $\Omega_\mathrm{m}$ profiles are broad and extend to
higher values than the posteriors. We find,
\begin{align}
S_8 &= 0.809^{+0.043}_{-0.022}
        \;\left(0.813\pm0.020\right), \nonumber\\
\Omega_\mathrm{m} &= 0.301^{+0.075}_{-0.083}
        \;\left(0.278^{+0.051}_{-0.042}\right),
\label{eq:constraints_shear}
\end{align}
where the first value is the profile best fit with its
$\Delta\chi^2 \leq 1$ interval and the value in parentheses is the marginal
posterior mean with its 68\% credible interval. 

Adding the compressed
Planck 2018 high-$\ell$ TT,TE,EE Gaussian likelihood in $\omega_\mathrm{b}$,
$n_\mathrm{s}$ and $h$ increases the constraining power and brings the profiles and posteriors into agreement,
\begin{align}
S_8 &= 0.808\pm 0.019
        \;\left(0.805 \pm 0.017\right), \nonumber\\
\Omega_\mathrm{m} &= 0.309^{+0.043}_{-0.036}
        \;\left(0.309^{+0.042}_{-0.035}\right).
\label{eq:constraints_shear_cmb}
\end{align}
In the noiseless synthetic test, the best-fit values of $n_\mathrm{s}$, $h$ and $\omega_\mathrm{b}$ jump between the edges of their parameter range along the profile (Fig.~\ref{fig:noiseless_data_profile}, A2). When all the unconstrained parameters were fixed, the profile and posterior agree closely (Fig.~\ref{fig:noiseless_data_profile}, B1). After adding the CMB likelihood, the same happens in Fig.~\ref{fig:desy3_plus_cmb}, where the best-fit parameters along the profile path no longer reach the parameter boundaries, bringing profiles and posteriors into agreement.
\subsection{Robustness to baryon feedback modeling}
\label{sec:baryons}
\begin{figure}
    \centering
    \includegraphics[width=1.0\linewidth]{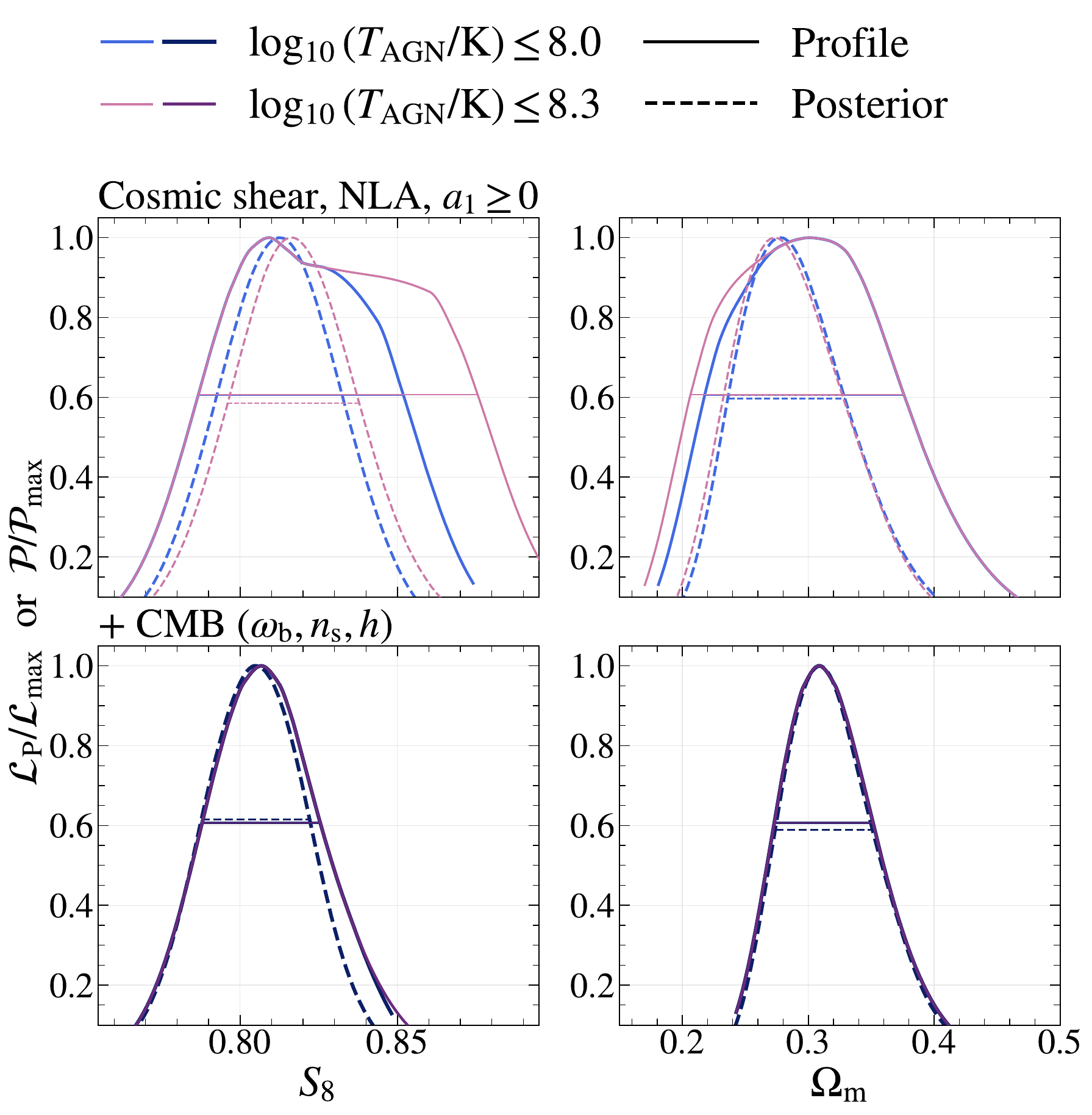}
    \caption{Profiles (solid) and posterior (dashed) comparison for two upper prior bounds on baryonic feedback strength. \emph{Top:} For cosmic shear alone, the degeneracy between $S_8$ and feedback makes the profile flat at high $S_8$, and the flat part extends with the upper bound. \emph{Bottom:} With the CMB likelihood, profiles and posteriors agree and do not depend on the upper bound on $\log_{10} (T_\mathrm{AGN}/\mathrm{K})$.} 
    \label{fig:desy3_tagn}
\end{figure}
The physical degeneracy between $S_8$ and baryonic feedback ($T_{\mathrm{AGN}}$) arises because both affect the matter power spectrum similarly on non-linear scales. In Fig.~\ref{fig:desy3_tagn}, we find that for cosmic shear alone, strong feedback is degenerate with $S_8$ in the high-$S_8$ tail. When the upper limit on $\log_{10} (T_\mathrm{AGN}/\mathrm{K})$ is extended from 8.0 to 8.3, the best fit remains unchanged, but the profile becomes effectively flat at high $S_8$, and its upper $1\sigma$ limit moves to higher values. The shift in the $S_8$ posterior is therefore driven by the $\log_{10} (T_\mathrm{AGN}/\mathrm{K})$ prior. The addition of the CMB likelihood removes this dependence, bringing the profiles and posteriors into agreement. The flatness in $S_8$ profile with high $\log_{10}(T_\mathrm{AGN}/\mathrm{K})$ in the top panel of Fig.~\ref{fig:desy3_tagn} is seen in the 2D profiles in Fig.~\ref{fig:desy3_tagn_2d}. For cosmic shear alone, the 2D contours rise towards higher $S_8$ as feedback is increased. The addition of the CMB likelihood breaks this degeneracy, making the contours nearly horizontal and $S_8$ constraints insensitive to the upper limit on $\log_{10}(T_\mathrm{AGN}/\mathrm{K})$. 

\subsection{Constraining baryonic feedback with cosmic shear on small scales and CMB}
\begin{figure}
    \centering
    \includegraphics[width=1.0\linewidth]{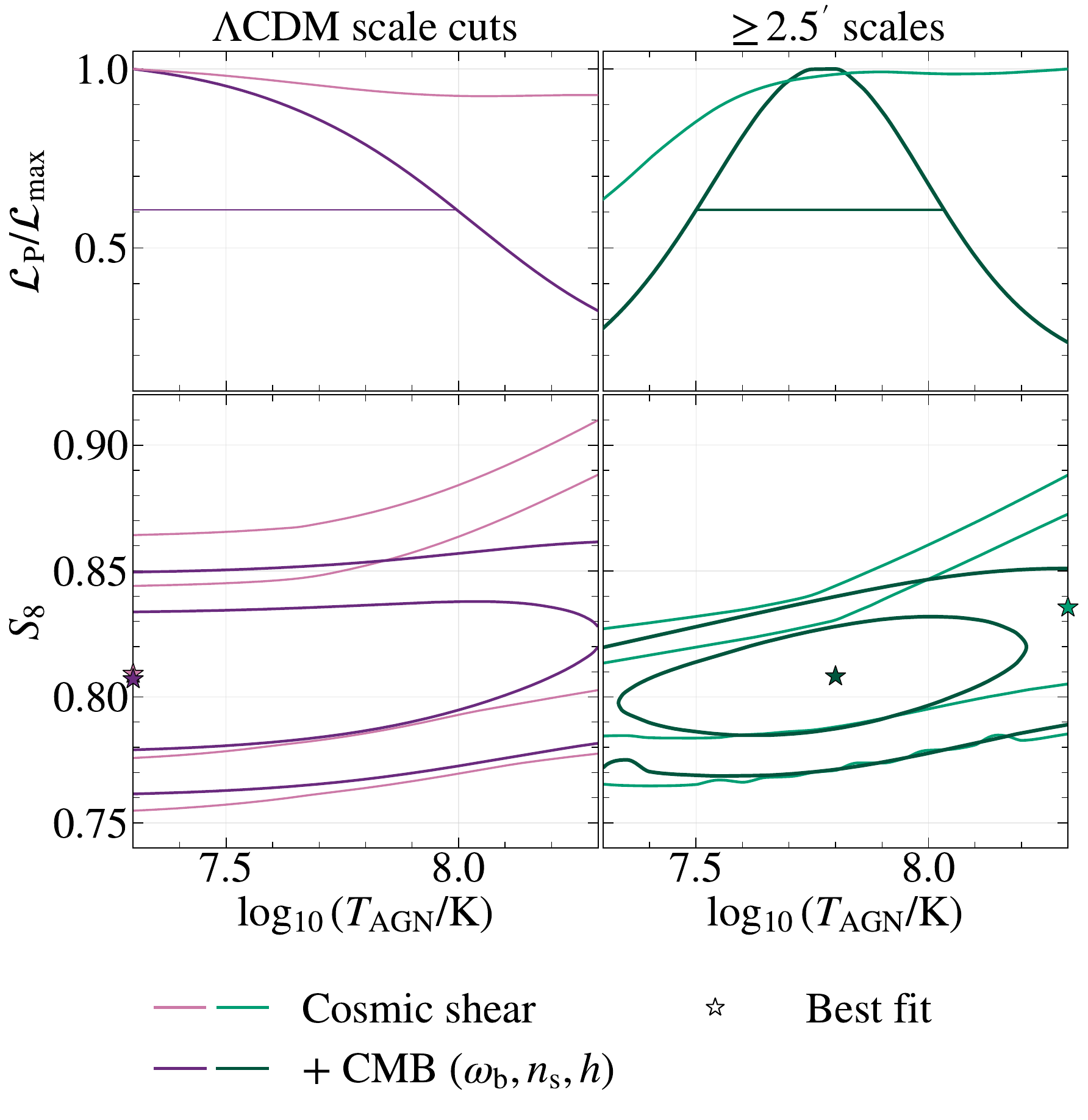}
    \caption{1D profiles of $\log_{10}(T_\mathrm{AGN}/\mathrm{K})$ (top) and 2D profiles in $\log_{10}(T_\mathrm{AGN}/\mathrm{K})$ and $S_8$ (bottom) for $\Lambda$CDM-optimized scale cuts (left) and all scales $\geq 2.5'$ (right), with the upper prior bound extended to $8.3$. Cosmic shear alone is unable to break high $S_8$-high feedback degeneracy. With small scales and the CMB, the profile prefers moderate feedback. The global best-fit $S_8$ value is unaffected by the choice of scale cuts.}
    \label{fig:desy3_tagn_2d}
\end{figure}
In Fig.~\ref{fig:desy3_tagn_2d}, $\Lambda$CDM-optimized scale cuts combined with the CMB give only an upper limit on $\log_{10}(T_\mathrm{AGN}/\mathrm{K})$. Including scales down to $2.5'$ closes the $1\sigma$ contour, and the profiles give,
\begin{align}
\log_{10}(T_\mathrm{AGN}/\mathrm{K})&\leq 8.0\hspace{1cm}\left(\Lambda\mathrm{CDM} \,\mathrm{scale\,cuts}\,+ \mathrm{CMB}\right)\,, \nonumber\\
\log_{10}(T_\mathrm{AGN}/\mathrm{K}) &= 7.80^{+0.23}_{-0.30}\;\left(\geq 2.5' \,\mathrm{scales}\,+ \mathrm{CMB}\right) .
\label{eq:tagn}
\end{align}
The best fit agrees with the calibrated BAHAMAS value, $\log_{10} (T_\mathrm{AGN}/\mathrm{K}) = 7.8$~\cite{McCarthy2017}, and the interval largely overlaps the range $7.6$--$8.0$ over which BAHAMAS reproduces the observed halo gas fractions. The data prefer moderate feedback, consistent with the upper limit obtained with $\Lambda$CDM-optimized scale cuts and the CMB. With the CMB likelihood, the global best-fit $S_8$ is unaffected by the choice of scales. The $\Lambda$CDM-optimized scale cuts therefore mitigate the impact of baryons without biasing $S_8$.
\begin{figure}
    \centering
    \includegraphics[width=1.0\linewidth]{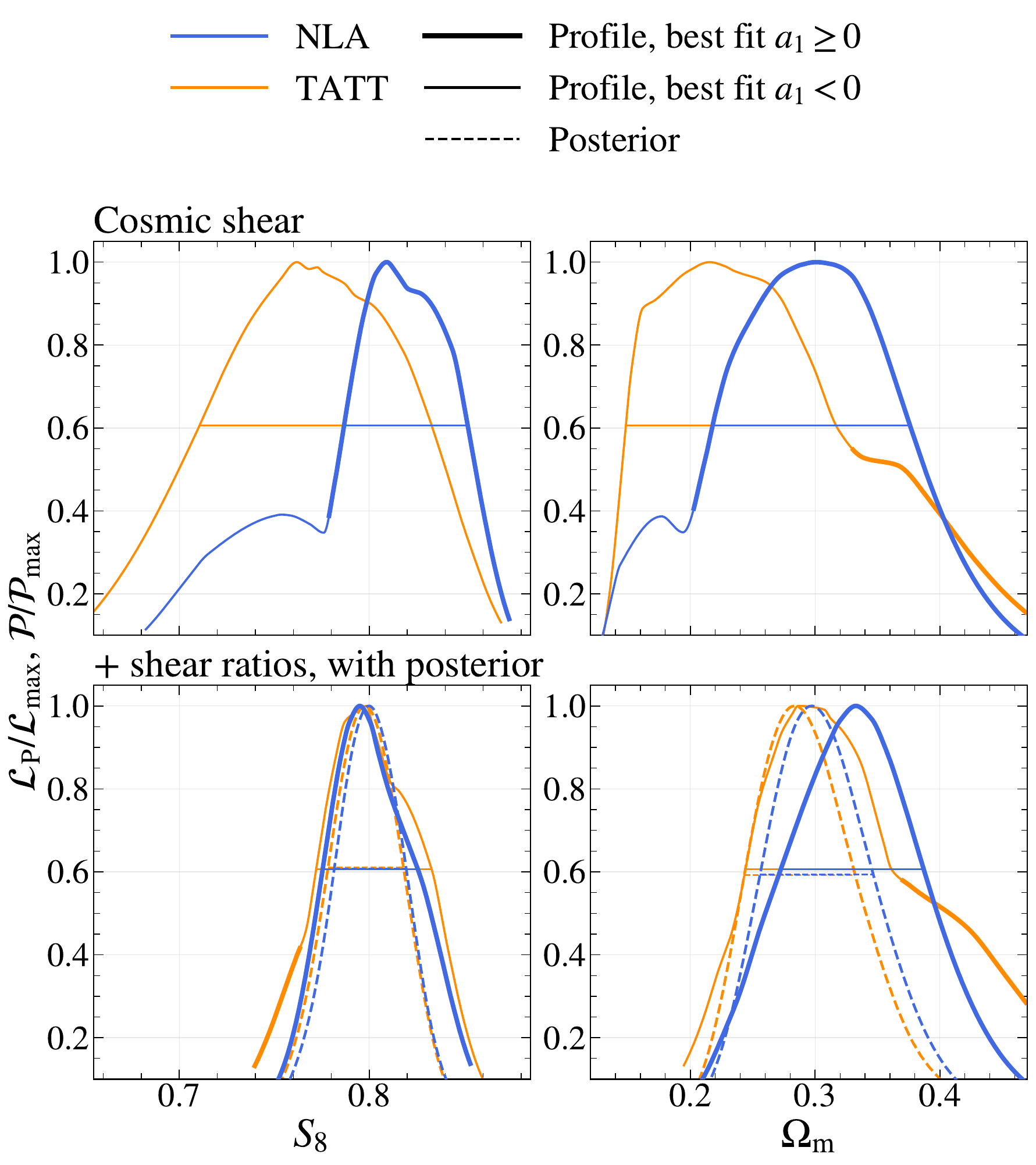}
    \caption{Profile comparison between \NLA{} and TATT models with $a_1$ in range $[-5,5]$. Thick lines mark points along the profile where the best fit has $a_1\geq0$, thin lines where it has $a_1<0$.
    \emph{Top:} for cosmic shear alone, the \NLA{} profile has a secondary peak on the $a_1<0$ branch at low $S_8$. The TATT best fit lies on the $a_1<0$ branch, and its profile peaks at lower $S_8$ and is much broader.
    \emph{Bottom:} DES Y3 shear ratios remove the \NLA{} $a_1<0$ branch and bring the \NLA{} and TATT $S_8$ profiles and the marginal posteriors (dashed) into close agreement.}
    \label{fig:desy3_nla_tatt}
\end{figure}

\subsection{Robustness to IA modeling}
\label{sec:ia_robustness}
The degeneracy between the IA parameters and $S_8$ arises because the IA terms $C_\mathrm{GI}$ and $C_\mathrm{II}$ in Eq.~(\ref{eq:Cl_EE_BB}) add to the lensing term, $C_\mathrm{GG}\propto S_8^2$. With our sign convention (Eq.~\ref{eq:nla_amp}), a negative alignment amplitude $a_1$ makes $C_\mathrm{GI}$ positive, so a lower $S_8$ can be compensated by a more negative $a_1$. Extending \NLA{} to TATT adds higher-order contributions to $C_\mathrm{GI}$ and $C_\mathrm{II}$, which open further degenerate directions in the space of $S_8$, $\Omega_\mathrm{m}$ and the IA parameters. The comparison between the \NLA{} and TATT models is given in Fig.~\ref{fig:desy3_nla_tatt}. For cosmic shear alone (top row), the \NLA{} best fit has $a_1\geq0$. The $a_1<0$ best-fit solution appears only at low $S_8$, outside the $1\sigma$ interval. The TATT best fit has $a_1<0$ and lies at lower $S_8$. Cosmic shear alone constrains the additional TATT parameters only weakly, resulting in TATT profiles being much less constrained than \NLA{} (as also seen for noiseless synthetic data in Fig.~\ref{fig:toy_noiseless_NLA_TATT}). Adding shear ratios removes the \NLA{} solutions with $a_1<0$ and brings the \NLA{} and TATT $S_8$ profiles and posteriors into close agreement. We investigate IA models in more detail with 2D profiles in Fig.~\ref{fig:desy3_lcdm_ia_free} (App.~\ref{app:ia}).

\section{Forecast for Roman (LSST wide) survey}
\label{sec:roman}
\begin{figure}[!t]
    \centering
    \includegraphics[width=0.95\linewidth]{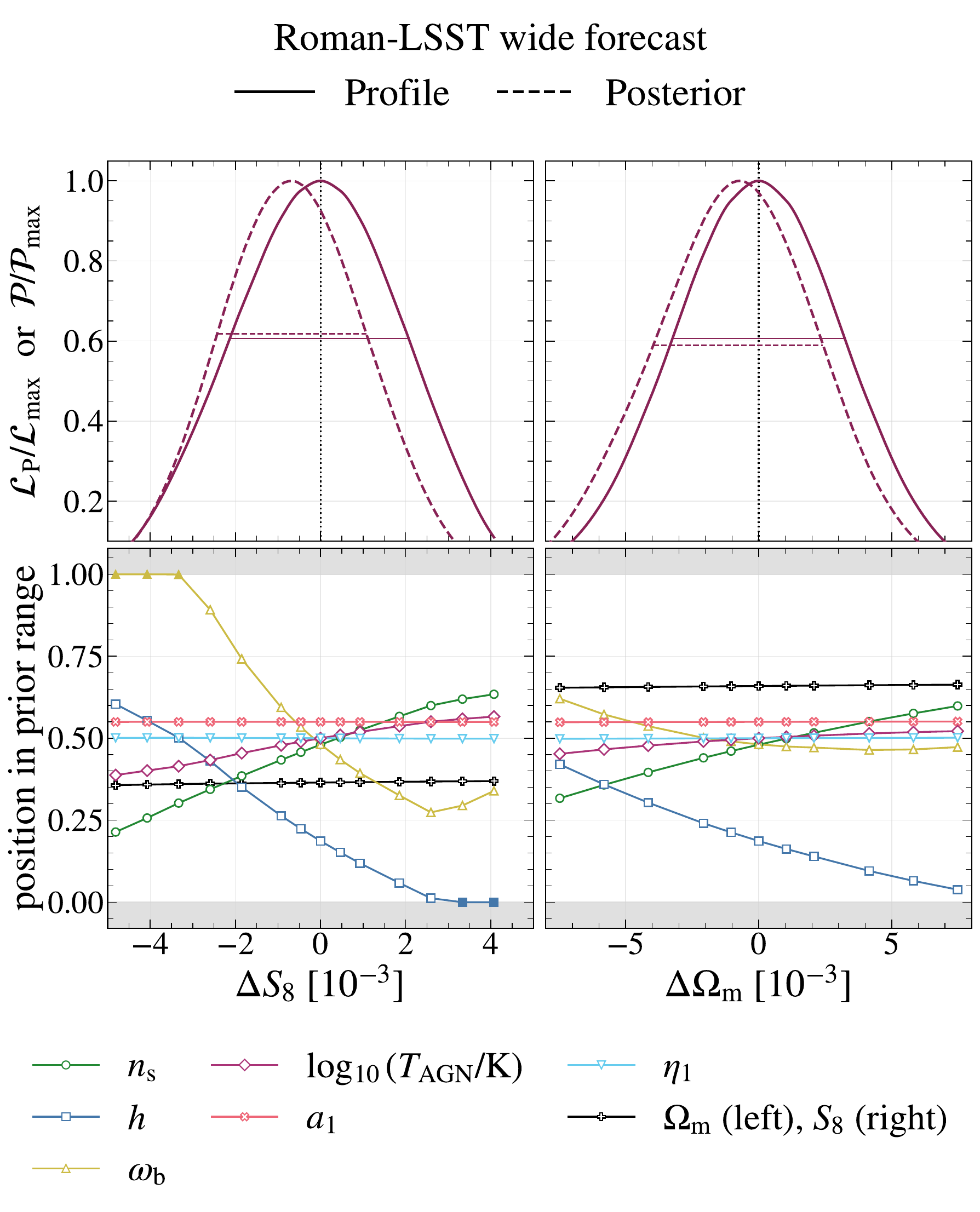}

    \caption{Forecast for a wide Roman survey over the LSST Y10 footprint. \emph{Top:} profile likelihoods (solid) and marginal posteriors (dashed), shown relative to the fiducial values (dotted line). Horizontal bars show the $1\sigma$ intervals.
    \emph{Bottom:} best-fit values of the other parameters along each profile, shown as their position within the prior range, as in Fig.~\ref{fig:noiseless_data_profile}. The profiles recover the input values. Unlike the DES~Y3-like case of Fig.~\ref{fig:noiseless_data_profile}, most parameters are well constrained within the parameter boundaries, and profiles and posteriors agree more closely. The slight bias in the posteriors is caused by an asymmetric prior in weakly constrained parameters $h$ and $\omega_\mathrm{b}$ around the true value.}
    \label{fig:romanwide}
\end{figure}

To test whether the differences between profile likelihoods and marginal posteriors persist at Stage-IV precision, we perform a forecast on noiseless cosmic-shear data for a wide-area Roman survey covering the LSST Y10 footprint at Roman depth~\cite{Eifler2021b}. We assume the surface density of source galaxies $n_\mathrm{eff} = 43\,\mathrm{arcmin}^{-2}$, split equally among ten tomographic bins, with shape noise $\sigma_\epsilon = 0.37$ (total). We compute $\xi_\pm$ in 15 logarithmic angular bins between $2.5'$ and $250'$ with a Gaussian covariance computed with \textsc{CosmoCov}~\cite{Fang2020}. The mock data vector is computed assuming the NLA IA model and no photometric-redshift or shear-calibration bias. In simulated LSST cosmic shear analyses, the bias in marginal posteriors is caused by asymmetric prior boundaries relative to true parameter values~\cite{Robertson2026}. Profile likelihood analyses demonstrate that this bias arises because the likelihood is maximized at the prior boundaries for poorly constrained parameters; asymmetric bounds truncate the high-likelihood volume non-uniformly, inducing a systematic shift in the marginal posterior away from the true values (see Fig.~\ref{fig:toy_prior} for a toy illustration). With higher statistical precision, we find that most parameters are well-constrained in the prior range, yielding similar constraining power for profile likelihoods and marginal posteriors (Fig.~\ref{fig:romanwide}).

\section{Conclusion}
We test the robustness of parameter inference by comparing Bayesian posteriors with frequentist profile likelihood for DES Y3 cosmic shear only and in combination with CMB. Our main findings are:
\begin{itemize}
    \item Cosmic shear is only sensitive to $S_8$ and $\Omega_\mathrm{m}$. Along the profile, the best-fit values of the remaining cosmological and nuisance parameters lie at the edges of their parameter ranges where prior volume is negligible. As a result, for DES Y3 cosmic shear alone, Bayesian posteriors are tighter than profile likelihoods.
    
    \item Cosmic-shear-only constraints on $S_8$ are driven by the parameter range of the baryonic feedback model ($\log_{10}(T_\mathrm{AGN}/\mathrm{K})$). Stronger feedback suppresses the small-scale power and can compensate for a higher $S_8$.
    
    \item Extended intrinsic alignment models like TATT with additional parameters open new degenerate directions compared to NLA and are weakly constrained by cosmic shear data alone. Once shear ratios are incorporated, both Bayesian and profile likelihood constraints come in agreement.
    
    \item External information from CMB likelihood on $\omega_\mathrm{b}$, $n_\mathrm{s}$, and $h$ helps constrain the shape of the matter power spectrum, bringing profiles and posteriors into agreement. Cosmic shear data on small scales with CMB favors moderate feedback, consistent with hydrodynamical simulations calibrated to observed gas fractions.

    \item For DES Y3 cosmic shear alone, the upper end of the $S_8$ profile depends on the assumed upper limit on the feedback strength (Sec.~\ref{sec:baryons}), which limits what cosmic shear alone can say about the $S_8$ tension. With the CMB likelihood, profile and posterior agree with each other and are consistent with the \textit{Planck} $S_8$ value, within $1.0\sigma$ and $1.3\sigma$, respectively.
      
    \item With the enhanced statistical precision of Stage-IV surveys, the constraining power of profile likelihoods is expected to improve significantly. Our forecast for Roman (LSST wide) survey indicates that profile likelihoods will deliver constraints comparable to those obtained from Bayesian posteriors.

\end{itemize}
Our results demonstrate that profile likelihoods provide an essential cross-check to assess the impact of parameter degeneracies and prior-volume effects on Bayesian constraints. For cosmic shear-only analyses, calibrating $1\sigma$ profile intervals requires larger sets of noisy realizations, which we leave to future work.

\begin{acknowledgements}
We thank Elisa Ferreira, Tim Eifler, Jiachuan Xu, and Yu-Hsiu Huang for insightful discussions. AG, EK, and VM were supported in part by the Roman Project Infrastructure Team ``Maximizing Cosmological Science with the Roman High Latitude Imaging Survey'' (NASA contracts 80NM0018D0004-80NM0024F0012), NASA ATP ``Modeling Large-Scale Structure - Interfacing Theory, Simulations, and ML- accelerated Inference'' (award 23-ATP23-94) and Department of Energy grant DE-SC0025993. During this work, we used large language model assistants (Claude and Gemini) for code translation (from C to Python), code optimization, figure styling suggestions, and language editing of author-written text. All AI-assisted code was rigorously tested and verified by the authors.
\end{acknowledgements}

\setlength{\textfloatsep}{10pt plus 2pt minus 2pt}
\setlength{\floatsep}{8pt plus 2pt minus 2pt}
\setlength{\intextsep}{8pt plus 2pt minus 2pt}
\setlength{\abovecaptionskip}{4pt}
\setlength{\belowcaptionskip}{2pt}

\section{Toy model: Bayesian vs frequentist inference}
\label{app:toy}
\begin{figure}
    \centering
    \includegraphics[width=1.0\linewidth]{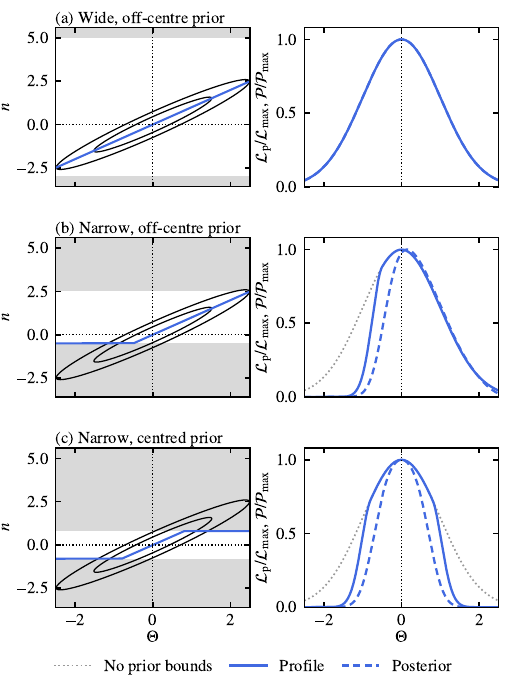}
    \caption{Effect of prior (parameter) boundaries on marginal posterior (profile likelihood).  We assume a Gaussian likelihood (Eq.~\ref{eq:toy_lkl}) with $\sigma=1$ and $\tilde{\sigma}=0.3$.}
    \label{fig:toy_prior}
\end{figure}
We compare Bayesian and frequentist parameter inference methods in a toy model with two parameters, $\Theta$: parameter of interest, and $n$: a nuisance parameter. We assume the likelihood to be a two-dimensional Gaussian centered at $\Theta=0=n$,
\begin{equation}
    -2\ln\mathcal{L}(\Theta,n) = \frac{\Theta^2}{\sigma^2} + \frac{(n-\Theta)^2}{\tilde{\sigma}^2}\,,
    \label{eq:toy_lkl}
\end{equation}
where the second term introduces correlation between the two parameters. At each $\Theta$, the profile likelihood $\mathcal{L}_\mathrm{p}$ takes the maximum of $\mathcal{L}$ over the allowed range of $n$. The posterior $\mathcal{P}$ integrates $\mathcal{L}$ over the same range. The posterior is a weighted average of the likelihood over the prior volume, whereas the profile is only sensitive to the likelihood peak in that volume. We compare the two statistics for different choices of nuisance parameter volume $V(\Theta)$ (see Eq.~\ref{eq:volume} for definition) in Fig.~\ref{fig:toy_prior}. This range is the prior range for the posterior and bounds the maximization for the profile. When $V(\Theta)$ spans the entire high-likelihood region, i.e., the priors are wide relative to the constraining power of the dataset, the two methods yield identical constraints. If $V(\Theta)$ is asymmetric around the maximum likelihood point, the posterior is biased and skewed toward regions of larger prior volume. The profile is also skewed but always peaks at the maximum likelihood point. If $V(\Theta)$ is symmetric, the posterior remains unbiased, but prior-volume effects shrink it relative to the profile.

\section{Analysis validation}

\subsection{Emulator and profile likelihood code}
\label{app:algorithm}
\begin{figure}[!t]
    \centering
    \includegraphics[width=1.0\linewidth]{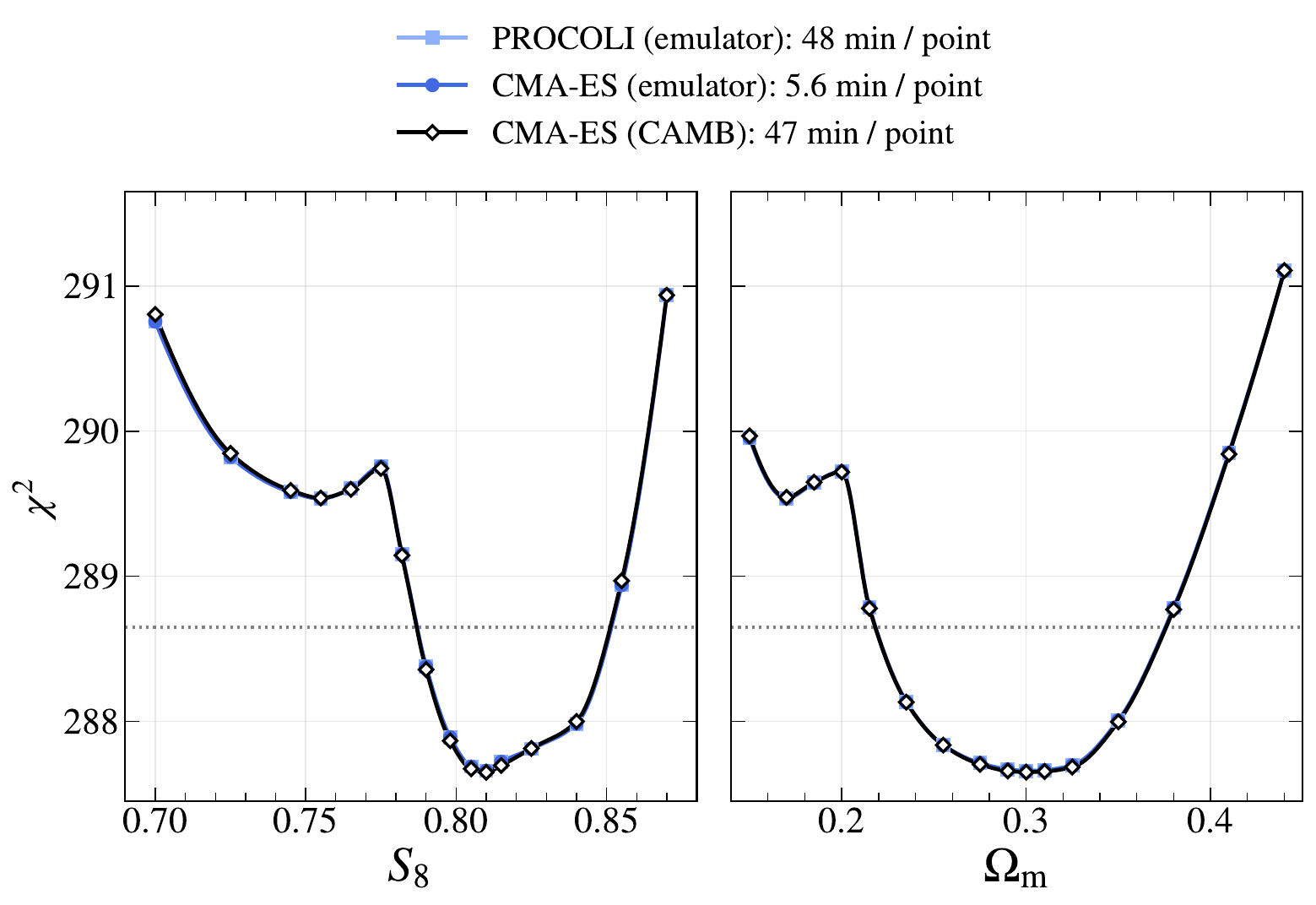}
    \caption{CAMB and emulator profiles run using CMA-ES and \textsc{PROCOLI} for NLA ($a_1\in[-5,5]$) are consistent. The absolute difference between $\chi^2$ values is $\leq 0.02$.}
    \label{fig:emulator_camb}
\end{figure}

\begin{figure*}
    \centering
    \includegraphics[width=1.0\linewidth]{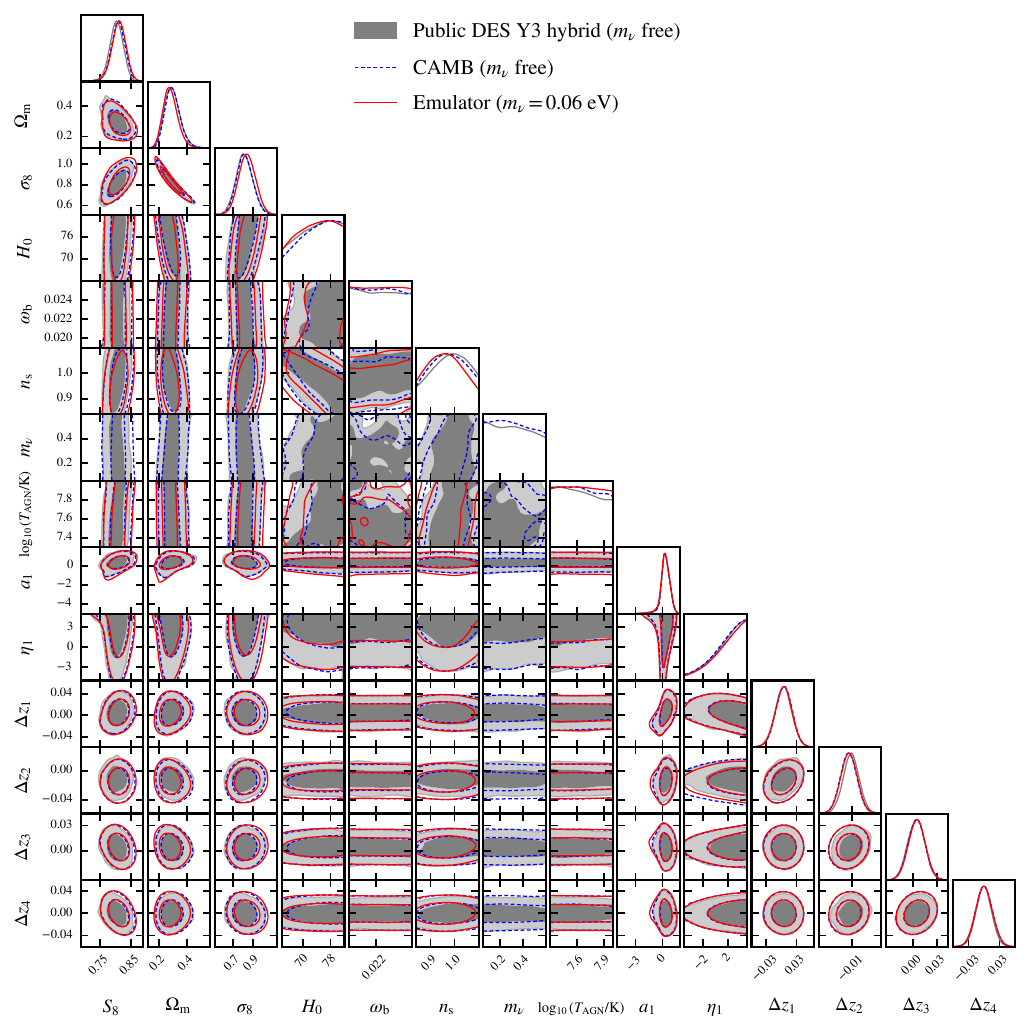}
    \caption{Posterior comparison with public DES-KiDS ($\Sigma m_\nu$ free), CAMB ($\Sigma m_\nu$ free), and KiDS Legacy emulator ($\Sigma m_\nu$ fixed) chains.}
    \label{fig:bmcmc_comparison}
\end{figure*}

The comparison of our analysis with public DES Y3 (Hybrid pipeline) is provided in Fig.~\ref{fig:bmcmc_comparison}. The profile likelihood code validation is provided in Fig.~\ref{fig:emulator_camb}.

\subsection{Noiseless and noisy synthetic data}
\label{app:validation}
\begin{figure}[!t]
    \centering
    \includegraphics[width=1.0\linewidth]{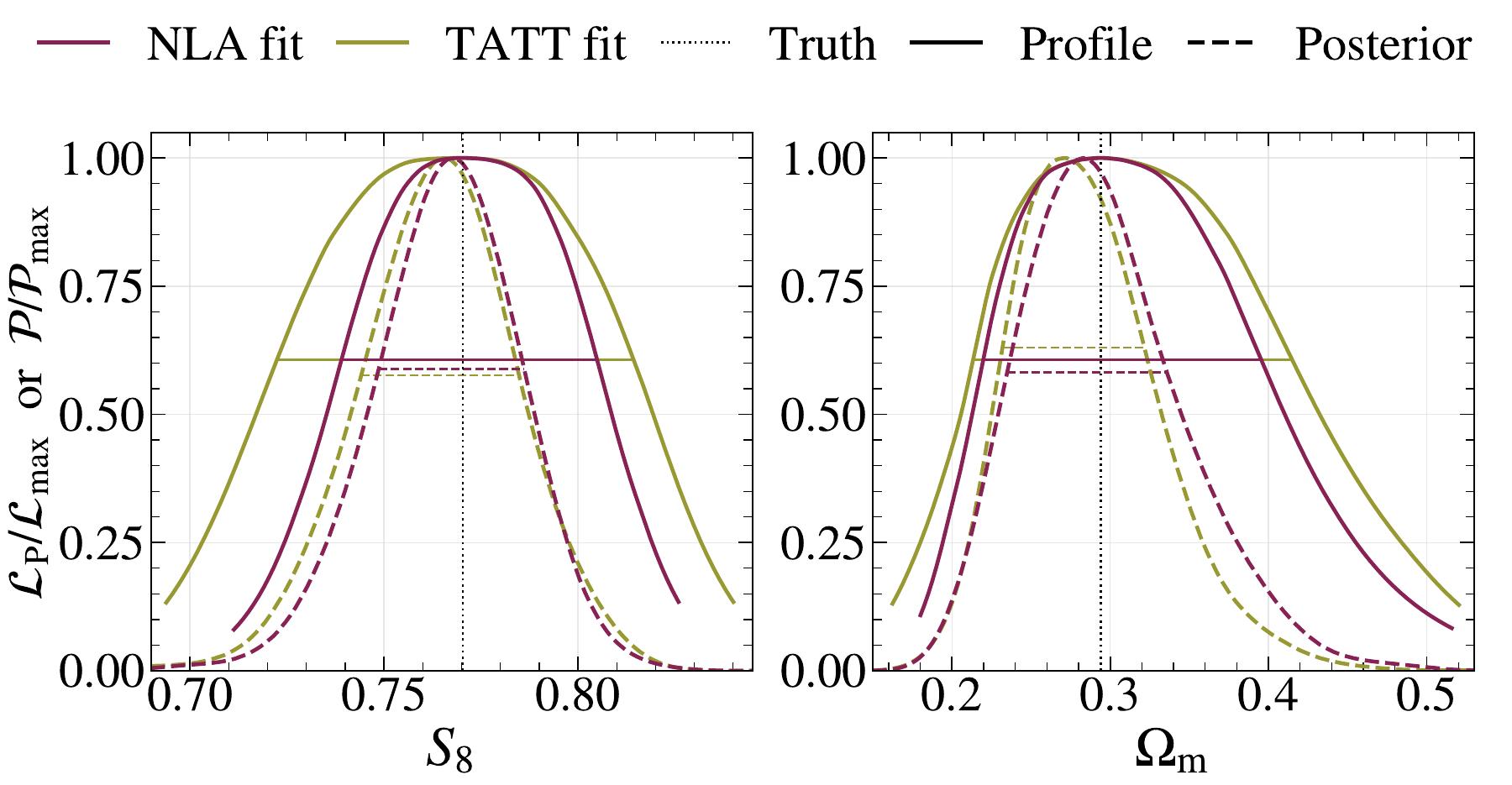}
    \caption{Same as the first row of Fig.~\ref{fig:noiseless_data_profile}, showing comparison of \NLA{} and TATT intrinsic-alignment models fitted to the same noiseless data vector, which was generated with \NLA{}. The three additional TATT parameters add extra freedom, which shows up in the profile becoming broader, but is largely suppressed in the posterior.}
    \label{fig:toy_noiseless_NLA_TATT}
\end{figure}

\begin{figure}[!t]
    \centering
    \includegraphics[width=0.9\linewidth]{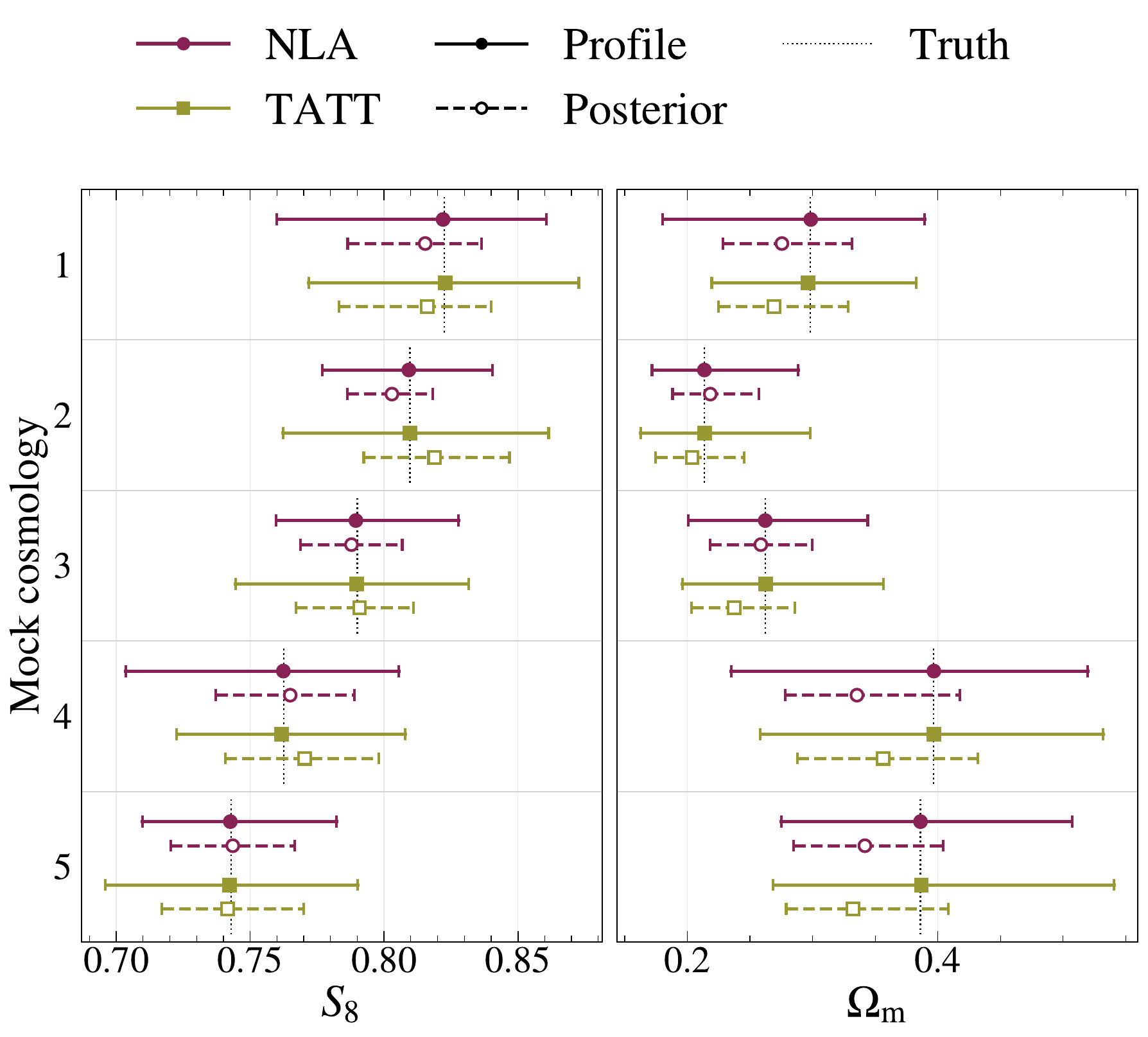}
    \caption{Validation on noiseless synthetic cosmic-shear data at five cosmologies drawn from a Latin hypercube. Each data vector is generated with NLA (wine) or TATT (olive) and fitted with the same IA model. For every cosmology and both IA models, the profile best fit recovers the input cosmology.}
    \label{fig:noiseless_cosmo}
\end{figure}

\begin{figure}[!tbp]
    \centering
    \includegraphics[width=\linewidth]{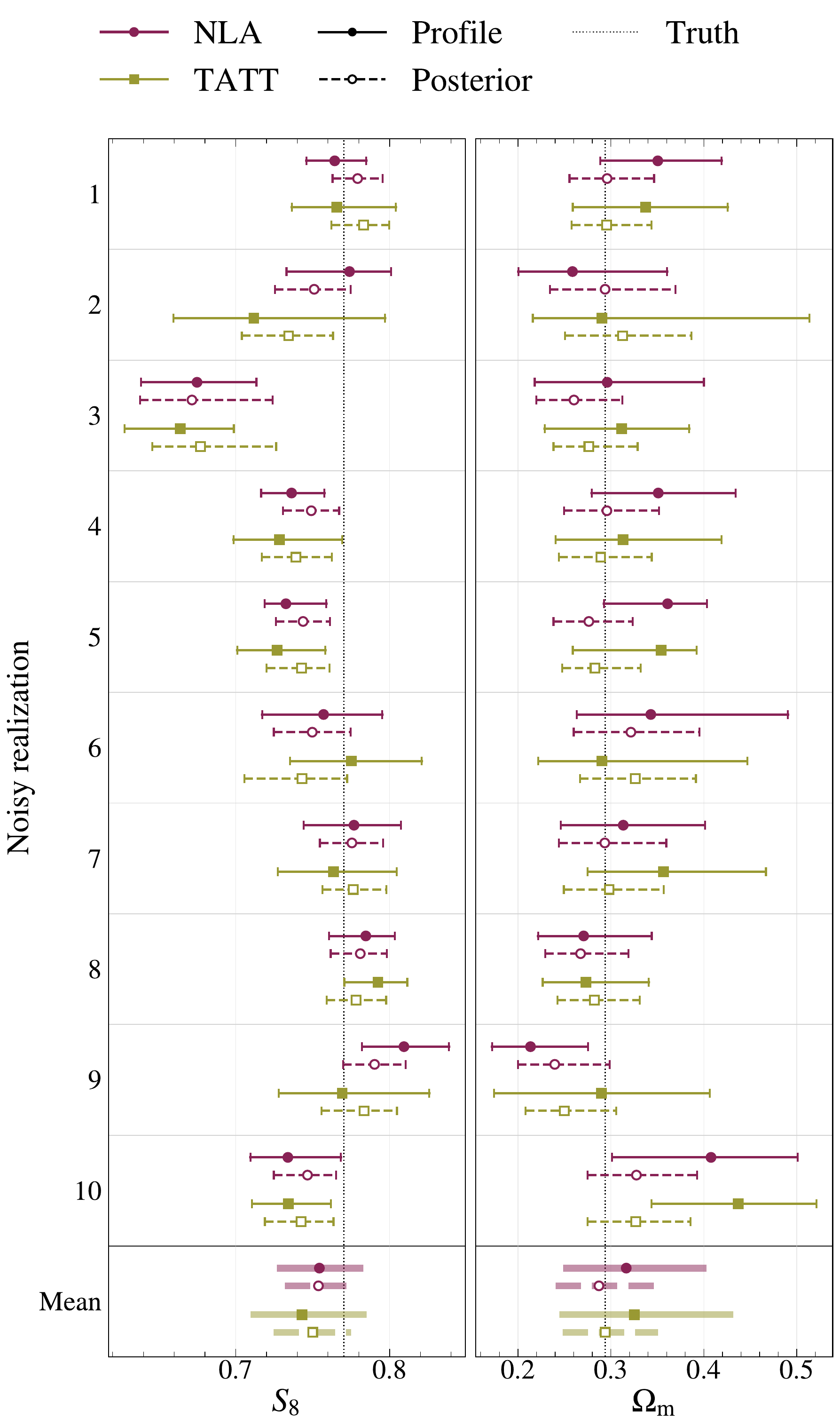}
    \caption{Validation on ten synthetic noisy realizations of cosmic shear data for NLA and TATT models. The interval constructed by averaging over ten noisy realizations contains the true value (black dotted line) at which noisy realizations were generated. More noise realizations are needed to calibrate the true coverage of the profile likelihood intervals.}
    \label{fig:noisy-data-profile}
\end{figure}

We validate our profile-likelihood pipeline on synthetic data vectors. We compare NLA and TATT models on noiseless datavector in Fig.~\ref{fig:toy_noiseless_NLA_TATT}. We first test noiseless realizations (to verify recovery of the input cosmology) in Fig.~\ref{fig:noiseless_cosmo} and then noisy realizations (to check the behavior of the frequentist intervals) in Fig.~\ref{fig:noisy-data-profile}.

\clearpage
\section{Extended analysis}
\label{app:mnu_ia}
\subsection{Freeing neutrino mass}
\label{app:mnu}
\begin{figure}[!tbp]
    \centering
    \includegraphics[width=1.0\linewidth]{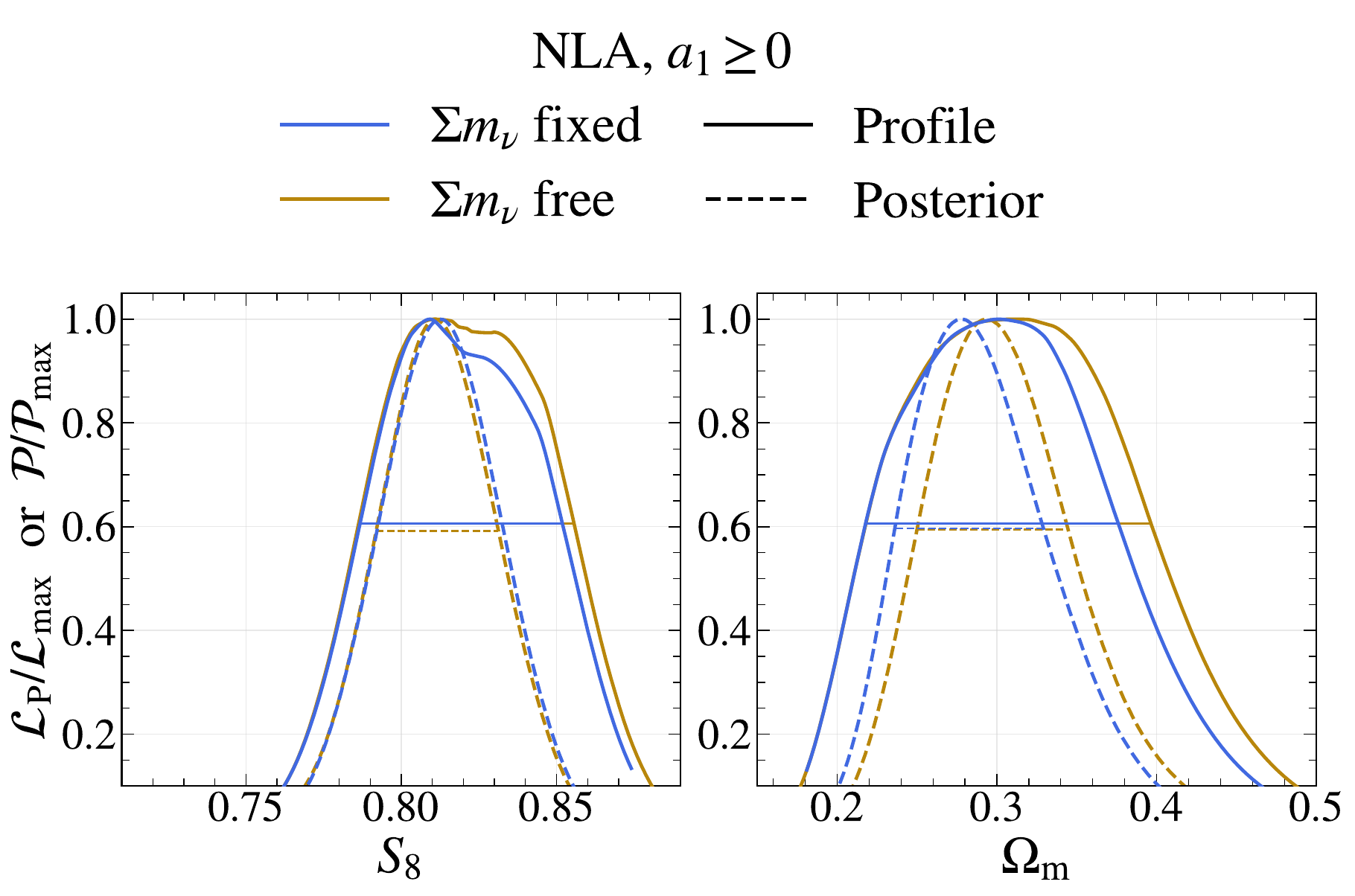}
    \caption{Allowing $\Sigma m_\nu$ to vary opens another degeneracy direction, skewing the profiles towards higher $S_8$ and $\Omega_\mathrm{m}$ values.}
    \label{fig:desy3_mnu}
\end{figure}
Our baseline analysis fixes $\Sigma m_\nu = 0.06$\,eV, whereas the Hybrid analysis of Ref.~\cite{DESKiDS2023} varies it with a uniform prior $\mathcal{U}[0.055, 0.6]$\,eV. Figure~\ref{fig:desy3_mnu} compares the two cases for cosmic shear alone with the \NLA{} model and $a_1 \geq 0$. Massive neutrinos suppress the matter power spectrum below their free-streaming scale, acting as an additional parameter that is degenerate with $S_8$ and $\Omega_\mathrm{m}$. Freeing $\Sigma m_\nu$ leaves $S_8$ almost unchanged but slightly skews $\Omega_\mathrm{m}$ profiles towards higher values.

\subsection{Intrinsic alignment: Removing \texorpdfstring{$a_1 \geq 0$}{a1 >= 0} restriction}
\label{app:ia}
\begin{figure}[!tbp]
    \centering
    \includegraphics[width=1.0\linewidth]{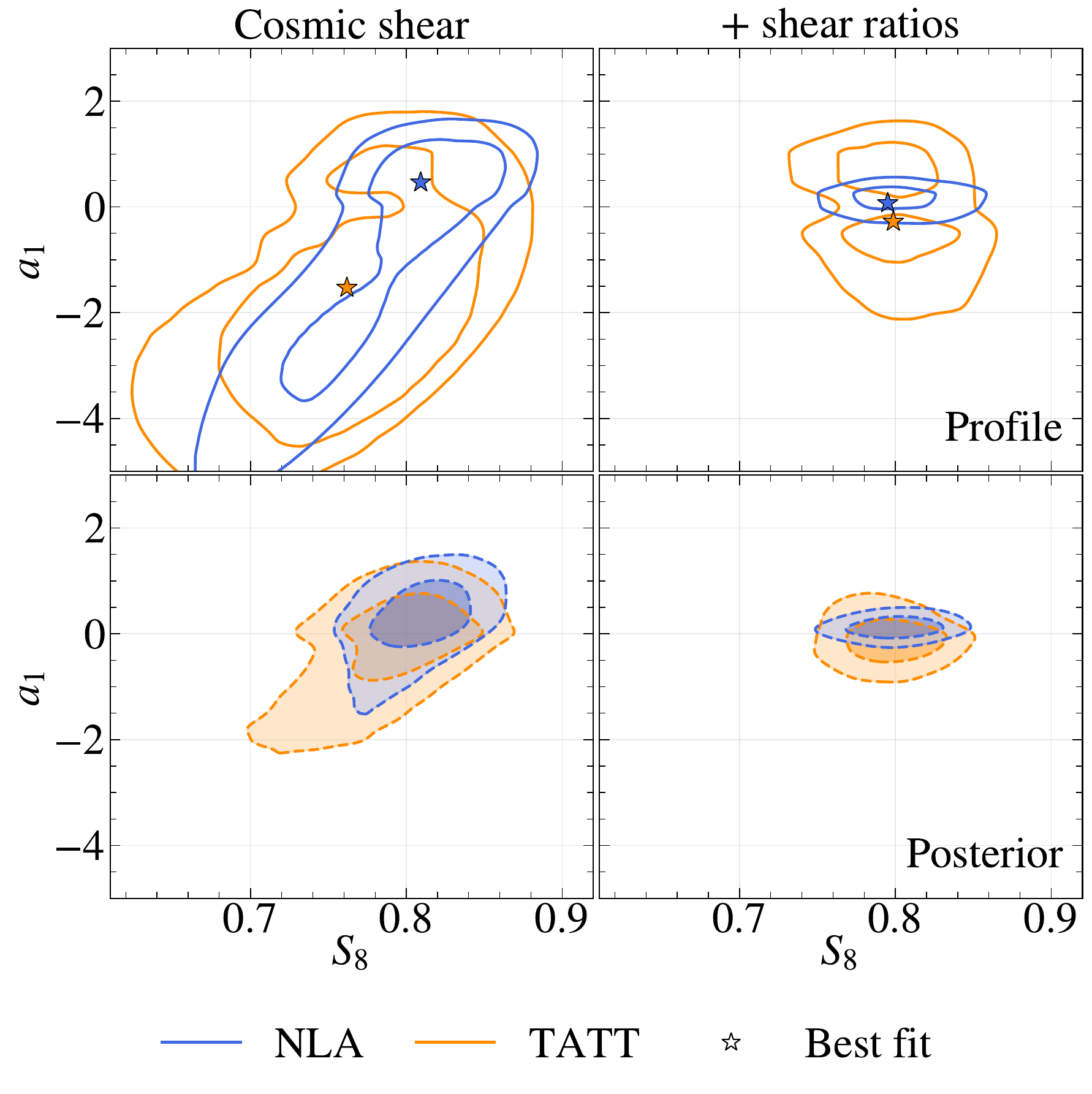}

    \caption{Both NLA and TATT profiles show bimodality in $a_1$. The best fit for NLA (TATT) lies in the $a_1\geq0$ ($a_1<0$) region. The addition of shear ratios helps constrain strong $a_1<0$ regions, bringing the posteriors and profiles for both models into agreement.}
    \label{fig:desy3_lcdm_ia_free}
\end{figure}

Figure~\ref{fig:desy3_lcdm_ia_free} compares the profile likelihood (top) with the posteriors (bottom) for the full range $a_1\in[-5,5]$. For cosmic shear alone, the profile is much broader than the posterior, by about a factor of 1.5 in $S_8$ for \NLA{} and 2 for TATT. The secondary maximum in profiles with large $a_1<0$ values is nearly as good as the $a_1\geq0$ region. However, $a_1<0$ regions in the profile prefer best-fit values of other weakly constrained cosmological and IA parameters close to the edges of their priors. The posterior gives little weight to prior edges and only explores the $a_1\geq0$ region. The shear ratios constrain $a_1$ independently of cosmic shear and remove the $a_1<0$ fits. For both models, the global best-fit $S_8$ agree, and the profile is only moderately broader than the posterior.

 \clearpage
\bibliography{ref.bib,y3kp.bib}
\bibliographystyle{unsrtads}
\end{document}